\documentclass[journal]{IEEEtran}
\usepackage{amsmath,amsfonts,amssymb}
\usepackage{algorithmic}
\usepackage{algorithm}
\usepackage{array}
\usepackage[caption=false,font=normalsize,labelfont=sf,textfont=sf]{subfig}
\usepackage{textcomp}
\usepackage{url}
\usepackage{verbatim}
\usepackage{graphicx}
\usepackage{overpic}  
\usepackage{cite}
\usepackage{booktabs}
\usepackage{doi}
\usepackage{multirow}
\newtheorem{definition}{\textbf{Definition}}

\newtheorem{remark}{\textbf{Remark}}
\DeclareUnicodeCharacter{207F}{$^{n}$}
\begin{document}
\title{Secure-CHG: A Comprehensive Framework for Robust and Fair Federated Learning via Hybrid Defense and Contribution-Aware Trust}

\author{Guanming~Che,
        Qiang~Wang$^{*}$,
        Jian~Xu,
        Fucai~Zhou
\thanks{The authors are with the Software College, Northeastern University, Shenyang 110013, China(e-mail: chegm@mails.neu.edu.cn; wangqiang1@mail.neu.edu.cn; xuj@mail.neu.edu.cn; fczhou@mail.neu.edu.cn).}
\thanks{This work was supported in part by the National Natural Science Foundation of China (Grants 62202090, 62173101, and 62372069), the Natural Science Foundation of Liaoning Province (Grant 2025-MS-046), and the Liaoning Collaboration Innovation Center for CSLE(Grant XTCX2024-015).}
\thanks{$^{*}$Corresponding author: Qiang Wang.}
}

\markboth{Journal of \LaTeX\ Class Files,~Vol.~14, No.~8, August~2021}%
{Shell \MakeLowercase{\textit{et al.}}: Secure-CHG}

\maketitle

\begin{abstract}
Federated Learning (FL) is highly susceptible to stealthy backdoor attacks, which aim to force a model into predicting an attacker-chosen target class for inputs containing a specific trigger. However, existing statistical defenses primarily focus on the early stages of model convergence. In this paper, we identify a fundamental vulnerability termed ``Late-stage Failure.'' We demonstrate that as the global model converges, decaying gradient norms render malicious and benign updates morphologically indistinguishable. This vanishing statistical variance effectively blinds traditional defenses, enabling adaptive adversaries to remain dormant and subsequently hijack the training process. To overcome these constraints, we propose Secure-CHG, a hybrid framework that pivots the defense paradigm from superficial morphological detection toward intrinsic semantic contribution verification. Secure-CHG employs an adaptive defense pipeline: a cascaded statistical filter stabilizes optimization during the early oscillatory phase, while a novel CHG-Shapley mechanism takes over during late-stage convergence. By leveraging sample hardness (i.e., local training loss) to project updates into a composite Hardness-Gradient space, it effectively amplifies adversarial semantic traces, enabling the isolation of stealthy attackers even as gradient norms vanish. Furthermore, we derive a closed-form solution for CHG-Shapley, facilitating low-complexity, retraining-free node valuation and trust-modulated aggregation. Extensive evaluations on CIFAR-10, MedMNIST, and NEU-SDDB demonstrate that Secure-CHG effectively mitigates Late-stage Failure. Specifically, it significantly suppresses advanced backdoor attacks, reducing their attack success rate by 2.3$\times$ and 2.0$\times$ relative to the mainstream Krum and Trimmed Mean baselines, respectively.
\end{abstract}


\section{Introduction}
\IEEEPARstart{F}{ederated} Learning (FL) has emerged as a promising decentralized learning paradigm that enables multiple clients to collaboratively train a global model without sharing their raw local data \cite{mcmahan2017communication, zhu2021federated}.
In a standard FL workflow, the server broadcasts the current global model, clients perform local training on their private datasets, and the server aggregates the uploaded updates to produce the global model for the next round.
This training pattern offers substantial privacy advantages and has made FL attractive for sensitive domains such as healthcare, finance, and industrial intelligence.
However, the distributed and partially untrusted nature of FL also exposes the training process to severe security threats, especially poisoning and backdoor attacks \cite{bagdasaryan2020backdoor, tolpegin2020data}.
Malicious clients can upload manipulated updates to degrade the global model or implant hidden triggers while appearing to participate normally \cite{shejwalkar2021manipulating}.
To defend against such threats, many robust aggregation methods, such as Krum \cite{blanchard2017machine}, Trimmed Mean \cite{yin2018byzantine}, and related statistical filters, attempt to identify abnormal clients based on geometric or statistical properties of their updates.
Since malicious updates exhibit explicit deviations in magnitude or direction, these defenses work well during the initial training stages.

In this paper, we reveal a critical theoretical deficiency in existing defense mechanisms that manifests as the model approaches convergence, a phenomenon we define as \textit{Late-stage Failure}.
As the global model gradually converges, the magnitudes of updates from both benign and malicious clients shrink toward zero and become increasingly similar.
Consequently, the absolute distances and relative variances between benign and malicious updates in the parameter space collapse.
As a result, distance-based and morphology-based defenses lose their discriminative power.
This creates a dangerous blind spot in the late training stage, allowing attackers to inject backdoors that bypass traditional checks without significantly compromising main-task performance.

\begin{figure}[!htbp]
    \centering   \includegraphics[width=\linewidth]{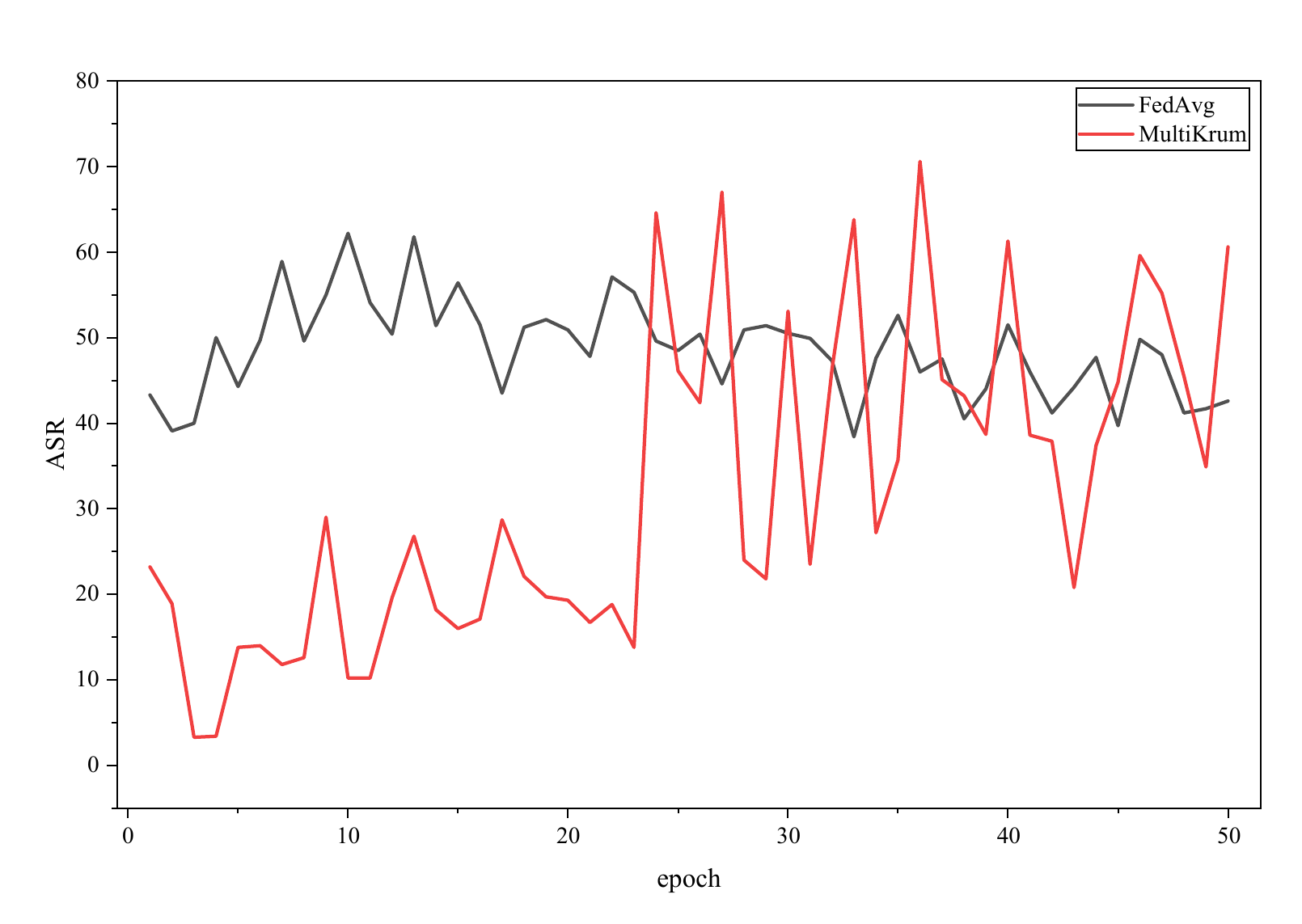}
    \caption{Illustration of Late-stage Failure: Defense capability collapses as the model converges.}
    \label{fig:fed_krum}
\end{figure}

As illustrated in Fig. \ref{fig:fed_krum}, we evaluate the ostensibly robust MultiKrum mechanism under a targeted label-flipping attack (e.g., misclassifying ``Cat'' as ``Dog'') with a 20\% malicious client ratio.
Our results demonstrate that MultiKrum's defense capability collapses abruptly after approximately 20 communication rounds.
In the subsequent 30 rounds, the average Attack Success Rate (ASR) surges to 41.97\%, indicating that nearly half of the adversarial updates successfully evade detection.
This precipitous decline in protection—a ``defense blind spot'' during the terminal phase of training—confirms that Late-stage Failure is not merely an implementation artifact of a single algorithm, but rather a systematic mismatch between the optimization dynamics of model convergence and the statistical assumptions underlying conventional anomaly detection.

In summary, the inherent Late-stage Failure exposes a critical vulnerability in traditional defenses.
As converging gradient norms render malicious and benign updates morphologically indistinguishable, the system inadvertently conflates stealthy adversaries with benign participants.
To transcend these vulnerabilities, we propose Secure-CHG, an intrinsically robust and computationally efficient FL governance framework.
Pioneering a full-lifecycle defense paradigm, Secure-CHG adaptively toggles between two governance phases based on the model's convergence state:

\begin{itemize}
    \item \textbf{Early-stage defense:} During the pre-convergence oscillatory period, a cascaded dual-filter (cosine similarity and FL-Defender \cite{jebreel2023fl}) precisely identifies and eliminates statistically anomalous gradients, aggressively purifying the initial training environment to stabilize the optimization trajectory.
    \item \textbf{Late-stage defense:}  To break the defensive deadlock caused by diminishing parameter variance, we introduce sample hardness (i.e., local training loss) as a semantic evaluation metric.
    Because poisoning tasks inherently conflict with primary optimization, malicious nodes exhibit persistently high hardness even as the global model stabilizes.
    By coupling this hardness metric with parameter updates, Secure-CHG effectively magnifies adversarial traces and isolates stealthy attackers precisely when statistical features fail.
    Crucially, it enables real-time, constant-complexity client valuation without external proxy datasets or costly model retraining.
\end{itemize}

Ultimately, an EMA-based long-term trust mechanism translates these real-time contributions into aggregation weights, forging a secure, equitable, and sustainable FL ecosystem.
\subsection{Our Contributions}
The main contributions can be summarized as follows:
\begin{enumerate}
    \item \textbf{Identifying and Mitigating Late-stage Failure in FL Defenses:} We expose the theoretical blind spot termed ``Late-stage Failure" where morphological convergence defeats distance-based statistical defenses. To counter this, we propose a semantic feature extraction method in the Hardness-Gradient space. By mapping sample hardness as weights into the update space, this approach amplifies the inherent conflict between poisoning and primary tasks, achieving precise isolation of stealthy attackers even as gradient norms vanish.
    \item \textbf{Proposing CHG-Shapley, an Efficient and Verification-Data-Free Valuation Algorithm:} Overcoming the exponential computational complexity and proxy-data reliance of traditional Shapley values, we derive a closed-form solution based on the composite Hardness-Gradient utility. This reduces valuation complexity to a constant level, providing a real-time, privacy-preserving assessment mechanism highly suitable for large-scale federated systems.
    \item \textbf{Designing Secure-CHG, an Adaptive Full-Lifecycle Framework:} Abandoning static defense strategies, we construct a dynamic pipeline that toggles based on convergence state. A cascaded statistical filter stabilizes early-stage oscillation, while a trust-modulated aggregation mechanism organically unifies late-stage malicious node defense with benign participant incentivization at the architectural level.
    \item \textbf{Extensive Empirical Validation Against Advanced Threats:} Evaluations across multi-modal datasets (CIFAR-10, MedMNIST, NEU-SDDB) demonstrate Secure-CHG's superiority. Compared to mainstream robust baselines (e.g., Krum, Trimmed Mean), it significantly suppresses late-stage advanced backdoors and label-flipping attacks, drastically reducing the Attack Success Rate (ASR) while preserving high global model utility across diverse heterogeneous scenarios.
\end{enumerate}

\subsection{Related Works}

\subsubsection{Defense Mechanisms in Federated Learning}

Existing FL defenses primarily fall into four categories, each facing inherent limitations. First, statistical and geometric approaches—--such as Krum \cite{blanchard2017machine}, RFA \cite{pillutla2022robust}, Trimmed Mean/Median \cite{yin2018byzantine}, FoolsGold \cite{samy2023mitigating}, and FLTrust \cite{cao2020fltrust}--—universally suffer from ``Late-stage Failure.'' As models converge, shrinking update norms cause statistical homogenization, rendering geometric discrepancies negligible and crippling detection sensitivity. Second, reinforcement learning and game-theoretic defenses \cite{yan2024byzantine, li2024meta} offer adaptability but incur exorbitant computational costs and unrealistically rely on clean validation sets for reward signals. Third, differential privacy-inspired methods, including \cite{zhang2022understanding, wang2025adaptive, xue2023differentially, li2024clients}, physically constrain update magnitudes. However, gradient clipping and high-variance noise fundamentally disrupt optimization, decelerating convergence and degrading model accuracy \cite{bagdasaryan2020backdoor}. Finally, defenses utilizing proxy datasets or server-side trust roots \cite{luo2024byzantine,sun2024byzantine, feng2025cgfl, zafar2025robust} overcome the lack of objective reference frames but paradoxically violate FL's core privacy principles and incur massive computational overhead due to domain gaps in synthetic data.
\subsubsection{Contribution Evaluation Mechanisms in Federated Learning}

Existing contribution evaluation mechanisms in FL primarily fall into three categories. First, Shapley Value (SV) methods \cite{wang2019measure, wang2022efficient,shapley1953value} provide a robust theoretical foundation. Despite lightweight approximations like auxiliary model fitting \cite{zhao2021efficient}, GTG-Shapley \cite{liu2022gtg}, and FedCon \cite{gao2024fedcon}, the prohibitive computational overhead remains an insurmountable bottleneck for massive networks or high-dimensional models\cite{cai2024chg}. Second, gradient consistency approaches utilize metrics like cosine similarity as an SV proxy to assess contributions efficiently \cite{zhang2022fedcos, zhang2024layer}. However, this geometric paradigm harbors a significant fairness flaw: it inherently rewards conformity. Minority clients with rare features naturally generate divergent gradients and are unjustly penalized as low-similarity outliers, failing to reflect their authentic value. Finally, proxy dataset-dependent evaluations quantify client value based on an independent validation set \cite{xie2019zeno, wang2024shapfed, apellaniz2025improving, lim2025fedeach}. These methods face a fundamental logical paradox: their efficacy relies entirely on an omniscient, unbiased, and centralized proxy dataset, which contradicts the decentralized premise of FL and severely curtails their practical viability.

\subsection{Paper Organization}
The rest of the paper is organized as follows. After presenting preliminaries in Section \ref{sec.2}, Sections \ref{sec.3}--\ref{sec:unified_trust} introduce our secure hybrid defense framework, covering serial filtering, CHG verification, and unified trust mechanisms. Finally, Section \ref{sec:experiments} discusses experimental evaluations, and Section \ref{sec:conclusion} concludes.

\section{Preliminaries}
\label{sec.2}
In this section, we first give some necessary definitions that will be used in the rest of the paper.
Table \ref{tab:notations} summarizes the notation used in our paper.
\begin{table}[!htbp]
    \centering
    \caption{Notations}
    \label{tab:notations}
    \renewcommand{\arraystretch}{1.2} 
    \begin{tabular}{ll}
        \toprule
        \textbf{Symbol} & \textbf{Description} \\
        \midrule
        $K$ & Total number of clients \\
        $\mathcal{K}$ & Set of clients \\
        $\eta$ & Learning rate \\
        $\epsilon$ & Small positive number (numerical stability/threshold) \\
        $\theta^t$ & Global model parameters at round $t$ \\
        $\theta_k^{t+1}$ & Local model parameters of client $k$ \\
        $\Delta\theta_k^t$ & Model update of client $k$ \\
        $\Delta\theta^t$ & Average of update vectors from all clients \\
        $\theta^{t+1}$ & Aggregated global model \\
        $D_k$ & Local dataset of client $k$ \\
        $n_k$ & Data volume of client $k$ \\
        $N$ & Set of all data points \\
        $N_{\text{total}}$ & Total data volume of all participants \\
        $i$ & Index of a single data point \\
        $L$ & Loss function \\
        $L_j$ & Loss of client $j$ \\
        $f(\theta)$ & Loss of model under parameters $\theta$ \\
        $f(i;\theta)$ & Loss of data point $i$ \\
        $\nabla f(\theta)$ & Gradient of the loss function \\
        $\nabla_{\text{last}}L_k$ & Last-layer gradient of client $k$ \\
        $U(S)$ & Utility function of subset $S$ \\
        $\phi_i(U)$ & Shapley value of data point $i$ \\
        $S$ & Subset of data points \\
        $h_i$ & Hardness of data point $i$ \\
        $C(S)$ & Proxy utility function \\
        $x_i$ & Gradient of data point $i$ \\
        $\theta_x$ & Parameters updated using gradient $x$ \\
        $\text{sim}(j,t)$ & Cosine similarity \\
        $\zeta_j$ & Instantaneous trust score \\
        $\beta$ & EMA decay factor \\
        $T_j^t$ & Long-term reputation of client $j$ \\
        $T^t$ & Average reputation score \\
        $\gamma$ & Relative contribution filtering factor \\
        $\lambda$ & Reputation modulation rate \\
        $C_k^t$ & Instantaneous contribution of client $k$ \\
        $W_k^t$ & List of safe nodes \\
        \bottomrule
    \end{tabular}
\end{table}

\subsection{Federated Learning}
Federated Learning (FL) enables collaborative model training while preserving local data privacy. At each round $t$, selected clients compute local model updates $\Delta\theta_k^t$ using their private data $D_k$. The central server then updates the global model by aggregating these local updates via weighted averaging (e.g., FedAvg): $\theta^{t+1} = \theta^t + \sum_{k=1}^{K} \frac{n_k}{N_{\text{total}}} \Delta\theta_k^t$, where $N_{\text{total}}$ is the total number of participating data samples.

\subsection{Poisoning Attacks in Federated Learning}
Within the federated paradigm, poisoning attacks inject malicious updates to subvert the global model, categorized by objectives (untargeted vs. targeted) and execution means (data vs. model poisoning). Untargeted attacks indiscriminately degrade overall convergence, whereas targeted attacks stealthily misclassify specific samples while preserving main-task performance. To achieve these goals, adversaries generate statistically camouflaged updates via two prominent strategies:
\begin{itemize}
    \item \textbf{Label Flipping (LF) Attack:} Attackers manipulate their dataset $D_k$ by changing the label of samples from class $a$ to class $b$.
    \item \textbf{Backdoor Attack (BA):} Attackers implant a specific pattern (trigger) into samples and assign a target label, causing the global model to output the target label for polluted samples. Recent advancements in these threats also include collusive attacks that optimize triggers to further evade detection \cite{lyu2024coba}.
\end{itemize}

\subsection{Shapley Value}
Originating from cooperative game theory, the Shapley Value is considered the ``gold standard'' for data valuation.
Ghorbani \& Zou \cite{ghorbani2019data} introduced Data Shapley, treating each data point (or client) as a ``player.'' The utility function $U(S)$ represents the model performance (e.g., accuracy) trained on a subset $S$. The Shapley value is defined as:
\begin{equation}
    \phi_i(U) = \sum_{S \subseteq \mathcal{K} \setminus \{i\}} \frac{|S|!(K-|S|-1)!}{K!} [U(S \cup \{i\}) - U(S)]
\end{equation}
This formulation computes the weighted marginal contribution of player $i$ across all possible coalitions. While it satisfies fairness axioms (efficiency, symmetry, etc.), its exact computation requires evaluating $2^K$ subsets, resulting in exponential complexity. This computational bottleneck severely limits its applicability in large-scale FL systems.

\section{The Secure-CHG Hybrid Defense Framework}
\label{sec.3}
In this section, we first give the problem statement, including the threat model and the core challenges. Then, we present the overall architecture of our proposed hybrid defense framework.

\subsection{Problem Statement}
\label{3.1}
\subsubsection{System Model}
\label{3.1.1}
Consider a Federated Learning system consisting of a central server and $K$ clients. The server maintains the global model parameters $\theta^t$, while clients train models on local datasets and upload the updates to the server.

Federated Learning proceeds in iterative rounds. The workflow for each round (round $t$) includes the following steps:
\begin{enumerate}
    \item \textbf{Broadcast:} The server broadcasts the current global model parameters $\theta^t$ to the selected clients.
    \item \textbf{Local Training:} Clients perform local training on private datasets to obtain local model updates.
    \item \textbf{Upload:} Clients upload round-specific training information to the server.
    \item \textbf{Aggregation \& Update:} The server performs detection, evaluation, and aggregation on the information uploaded by clients to generate the new global model $\theta^{t+1}$.
\end{enumerate}

To support defense and contribution evaluation mechanisms, clients are required to upload the following three types of information to the server in each training round:
\begin{itemize}
    \item \textbf{Local Model Update $\Delta\theta_k$:} Captures the parameter update for the $k$-th client.
    \item \textbf{Last-Layer Gradient $\nabla^{last}_k$:} Extracted from the last layer of the neural network, used to characterize model behavior.
    \item \textbf{Sample Hardness Scalar $h_k$:} Quantifies the difficulty of fitting the current global model on the client's local data.
\end{itemize}
The server performs defense and aggregation operations solely based on the uploaded information described above, without accessing the clients' raw training data throughout the process.
\subsubsection{Problem Formulation and Assumptions}
To rigorously analyze the phenomenon of Late-stage Failure, we introduce the following assumptions. These assumptions characterize the typical operating environment of FL during the model convergence phase without loss of generality.

\textbf{Assumption 1 (Convergence Bounds):}
In the late stage of training, the global model parameters $\theta$ enter a stable convergence region. Consequently, the $L_2$-norms of local model updates from all clients decay as training progresses. Specifically, there exists a sequence of positive scalars $\epsilon_t$ such that:
\begin{equation}
\|\Delta\theta_k^t\| \le \epsilon_t, \quad \text{where } \lim_{t \to \infty} \epsilon_t = 0,
\end{equation}
holds for any client $k$.

\textbf{Assumption 2 (Benign Update Distribution):}
During the convergence phase, model updates from honest clients are modeled as zero-mean random perturbations satisfying a bounded second moment condition:
\begin{equation}
\mathbb{E}[\Delta\theta_h^t] = 0, \quad \mathbb{E}[\|\Delta\theta_h^t\|^2] \le \sigma^2.
\end{equation}

\textbf{Assumption 3 (Adversarial Constraints):}
To evade detection in the late stage, malicious clients are constrained to maintain stealth. Their submitted updates are bounded in magnitude to the same order as the perturbations of honest clients:
\begin{equation}
\|\Delta\theta_a^t\| \le c \cdot \epsilon_t,
\end{equation}
where $c$ is a constant scaling factor.

\textbf{Assumption 4 (Zero-Trust Server):}
The server possesses no auxiliary clean validation data or trusted proxy models. All defense and evaluation decisions rely exclusively on the uploaded model updates and their derived features.

\subsubsection{Threat Model}
We adopt a time-varying Byzantine threat model. Let $\mathcal{K}$ denote the set of participating clients, where an unknown subset $\mathcal{A} \subset \mathcal{K}$ is controlled by an adversary, satisfying $|\mathcal{A}| \le f$ and $f < |\mathcal{K}|/2$. Clients outside $\mathcal{A}$ (i.e., $\mathcal{K} \setminus \mathcal{A}$) are considered honest. The adversary fully controls their local datasets, training processes, and uploaded updates. At the communication level, malicious clients follow the FL protocol and do not tamper with message passing or interrupt the training process. However, adversaries may pursue multiple objectives during training, including degrading global model performance or manipulating contribution assessment and incentive distribution.

\paragraph{Early Stage (Destructive Redirection and Convergence Stagnation)}
In the early stages of training, the global model remains far from convergence, and client updates typically exhibit large magnitudes and high variability. During this phase, malicious clients actively disrupt the training process by uploading updates designed to induce instability or oscillation in the global optimization trajectory. Such behaviors aim to delay convergence and destabilize optimization.

\paragraph{Late Stage (Covert Persistence and Incentive Theft)}
As training progresses and the global model gradually converges, the magnitude of client updates decays toward zero. In this stage, the adversary's goal shifts from explicit destruction to \textit{covert persistence}. This means the adversary maintains attack behaviors that minimize impact on the main task performance while attempting to manipulate contribution assessment and incentive distribution (i.e., \textit{Incentive Theft}). 

\begin{definition}[Late-stage Failure]
    \label{def:1}
    Let the Federated Learning system contain a client set $\mathcal{N} = \mathcal{B} \cup \mathcal{M}$. In training round $t$, client $k$ uploads model update vector $\Delta\theta_k^t$. When the global model $\theta^t$ tends to converge ($t \to \infty$), the system enters a failure state satisfying the following condition: as the model converges, to fine-tune the model, the $L_2$-norms of all client update vectors approach zero:
\begin{equation}
\lim_{t \to \infty} \|\Delta\theta_k^t\|_2 = 0, \forall k \in \mathcal{N}
\end{equation}
\end{definition}

\paragraph{Adaptive Strategic Dormancy}
In addition to the two-stage attack behaviors described above, we further consider a \textit{White-box Adaptive Adversary} who has full knowledge of the defense and trust update mechanisms. Such an adversary may employ a strategy of \textit{Adaptive Strategic Dormancy}. Specifically, in the early stages of training, the adversary intentionally behaves benignly. This strategy allows the adversary to accumulate substantial historical trust through the system's trust update mechanism (e.g., Exponential Moving Average). Once the global model enters the late convergence stage and sufficient trust has been established, the adversary adaptively switches to malicious behavior, leveraging the inertia of accumulated historical trust to evade immediate detection.

\subsection{Overall Framework Architecture}
\label{sec:3.2}
To address these problems formalized in Section \ref{3.1}, we propose the Secure-CHG framework, which abandons static defense mechanisms in favor of a \textit{Dynamic Governance Pipeline} that is tightly coupled with the model's convergence state. The core philosophy of the framework is to employ high-precision statistical filtering during the early training phase to navigate through the oscillation period rapidly. Subsequently, upon detecting that the model has entered the convergence phase (the potential ``Late-stage Failure'' period), the system seamlessly transitions to a contribution verification mode based on sample hardness.

\begin{figure*}[htbp]
\centering
\includegraphics[width=0.8\linewidth]{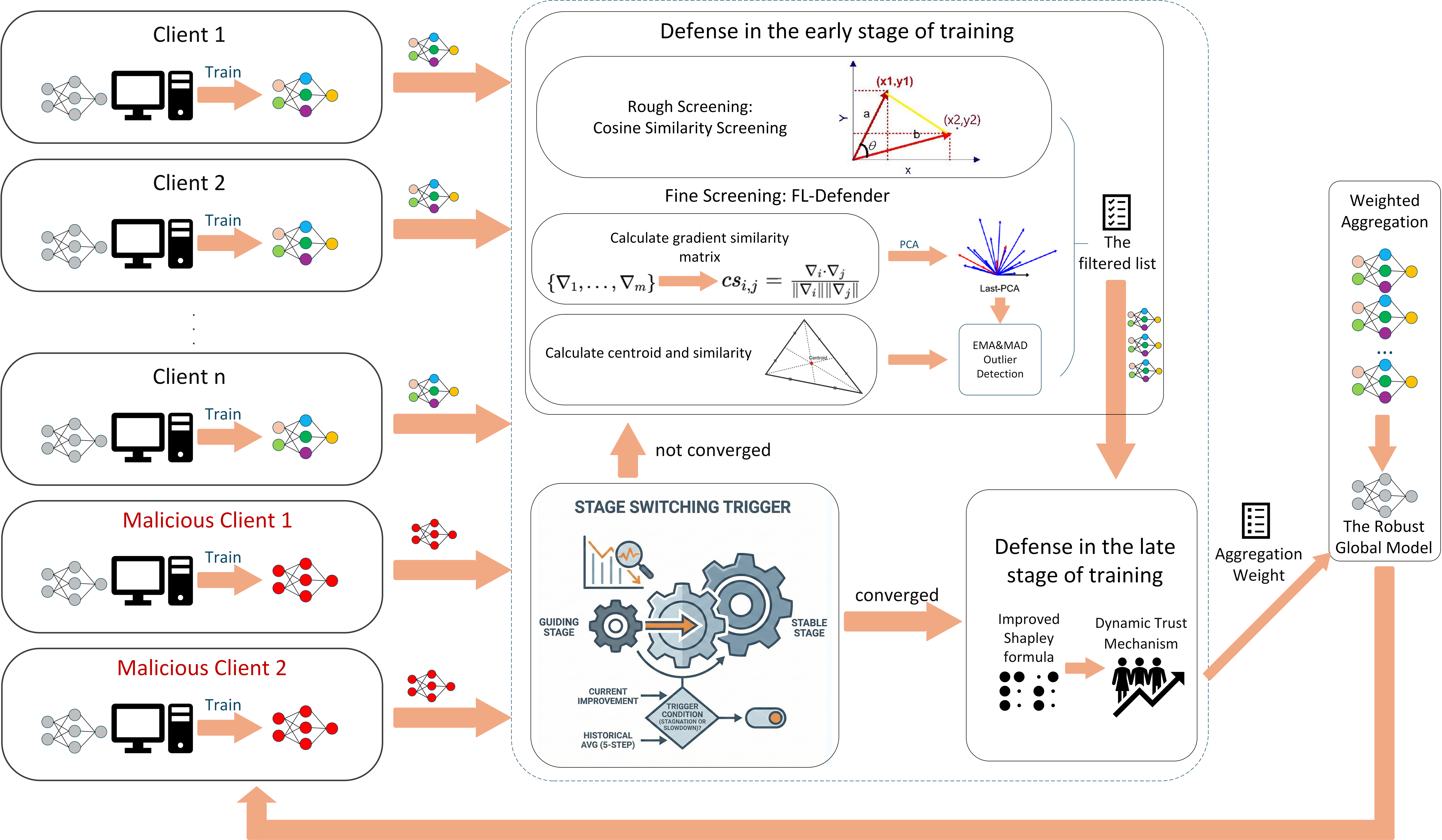}
\caption{The overall architecture of the Secure-CHG framework. The system features a dynamic governance pipeline that adapts to the convergence state, switching between ``Early-stage Defense'' and ``Late-stage Defense'' mechanisms.}
\label{fig:framework}
\end{figure*}

\subsubsection{Client Layer: Local Training and Feature Extraction}
Functioning as the edge computing nodes of the system, each client $k$ performs two primary tasks in every round $t$: local model optimization and the extraction of defense-related features. To facilitate the subsequent hybrid defense and contribution assessment, the client uploads a tuple $M_k^t = \{\Delta\theta_k^t, \nabla_{\text{last}}L_k, h_k\}$ to the server, where $\Delta\theta_k^t$ represents the model update, $\nabla_{\text{last}}L_k$ denotes the gradient of the last layer, and $h_k$ indicates the sample hardness.

\subsubsection{Server Layer: Convergence-Aware Dynamic Governance}
The server acts as the central decision-making unit. Its processing workflow consists of three serially connected stages: State Monitoring \& Routing, Dual-mode Defense, and Trust Aggregation.

\paragraph{State Monitoring \& Routing}
This module serves as the control center of the architecture. Prior to each aggregation round, the server assesses the physical stage of the global model based on the historical trend of the global loss. The system triggers a state transition by monitoring the ratio of the instantaneous loss reduction rate ($I_{\text{current}}$) to the historical average rate ($I_{\text{avg}}$). The specific logic is detailed in Algorithm \ref{alg:check_stage_switch}.

\begin{algorithm}
\caption{Adaptive Stage Switching Logic}
\label{alg:check_stage_switch}
\begin{algorithmic}[1]
\REQUIRE History of global loss values $L$, Current epoch index $t$, Boolean flag $is\_converged$ (False=Early, True=Late)
\STATE \textbf{Parameters:} $window = 5$ (Smoothing window size), $\epsilon = 10^{-6}$ (Stagnation threshold), $\gamma = 0.5$ (Slowdown ratio threshold), $\delta = 0.1$ (Reset threshold)
\STATE $I_{\text{current}} = L[t-1] - L[t]$ \COMMENT{Calculate instant improvement}
\STATE $I_{\text{avg}} = (L[t-window-1] - L[t-1]) / window$ \COMMENT{Calculate smoothed historical improvement}

\IF{$is\_converged == \text{False}$}
    \IF{$(I_{\text{avg}} \le \epsilon)$ \OR $(I_{\text{current}} < \gamma \cdot I_{\text{avg}})$}
        \STATE $new\_state = \text{True}$ \COMMENT{Switch to Late Stage (Contribution Verification)}
    \ELSE
        \STATE $new\_state = \text{False}$ \COMMENT{Remain in Early Stage}
    \ENDIF
\ELSE
    \IF{$I_{\text{current}} < -\delta$}
        \STATE $new\_state = \text{False}$ \COMMENT{Reset to Early Stage (Defense against collapse)}
    \ELSE
        \STATE $new\_state = \text{True}$ \COMMENT{Lock in Late Stage (Maintain stability)}
    \ENDIF
\ENDIF
\RETURN $new\_state$
\end{algorithmic}
\end{algorithm}

Based on the status flag $new\_state$ returned by Algorithm \ref{alg:check_stage_switch}, the system dynamically routes the data stream to distinct defense pathways:

\begin{itemize}
    \item \textbf{Early Stage ($new\_state = \text{False}$):} During unconverged loss oscillations, the defense prioritizes purifying the optimization environment. Data flows to \textbf{Path A (Serial Statistical Filtering $\to$ CHG Contribution Verification)} to eliminate high-amplitude outliers and accelerate convergence.

    \item \textbf{Late Stage ($new\_state = \text{True}$):} Triggered when convergence conditions ($I_{\text{avg}} \le \epsilon$ or $I_{\text{current}} < \gamma \cdot I_{\text{avg}}$) are met. As gradient norms decay, statistical features fail (Late-stage Failure). The system shifts to \textbf{Path B (Direct CHG Contribution Verification)}, bypassing statistical filters to identify covert malicious nodes at the semantic level via CHG-Shapley.

    \item \textbf{Hysteresis \& Anti-Jittering:} To prevent frequent state switching (``jittering'') caused by local SGD oscillations, the Late Stage state is locked once triggered. The system reverts to the Early Stage only if a severe loss rebound ($I_{\text{current}} < -\delta$) is detected.
\end{itemize}

\paragraph{Dual-Mode Defense \& Evaluation}
Depending on the routing decision, the system activates the corresponding processing logic:

\textbf{Module 1: Serial Statistical Filtering} \\
This module is strategically positioned for the \textbf{Early Stage} of Federated Learning. During this phase, the un-converged global model is highly susceptible to malicious gradients that can severely deviate the optimization direction. To ensure system integrity and balance computational feasibility, we employ a \textbf{Serial Statistical Filtering} approach, progressing from ``Macro Direction'' to ``Micro Structure'':
\begin{itemize}
    \item \textbf{Pass 1: Coarse-grained Macro Screening.} Using the full parameter update vector $\Delta\theta$, the system calculates cosine similarity to rapidly discard updates that exhibit negative similarity, indicating redirection attacks that deviate from the group consensus.
    \item \textbf{Pass 2: Fine-grained Micro Screening.} For updates passing the initial screening, the FL-Defender module analyzes the principal component distribution of the last-layer gradients $\nabla_{\text{last}}L$, filtering out nodes with anomalous micro-structures.
\end{itemize}
\textbf{Output:} A filtered list of secure nodes $W_k^t$. The decision state $w_k^t$ for client $k$ in round $t$ is defined as:
\begin{equation}
w_k^t = \begin{cases} 
1, & \text{if } \text{sim}(k,t) \ge 0 \text{ AND } H_k^t > Q1_0^t \\
0, & \text{otherwise (identified as malicious)}
\end{cases}
\end{equation}
If $w_k^t = 0$, the node is assigned a zero weight for the current round.

\textbf{Module 2: CHG Contribution Verification} \\
This module performs distinct core functions depending on the training stage:
\begin{itemize}
    \item \textbf{Early Stage (Path A - Serial Mode):} Functions as a secondary defense layer. It performs fine-grained value quantification on nodes that survived Module 1, preventing sophisticated attackers from accumulating unearned credit. Nodes rejected by Module 1 ($w_k^t=0$) are skipped, and their contribution is set to zero ($C_k^t=0$).
    \item \textbf{Late Stage (Path B - Direct Mode):} Functions as the primary line of defense. When gradient norm decay renders traditional statistical features ineffective, this module takes over using the CHG-Shapley mechanism. It identifies and eliminates malicious nodes that are statistically indistinguishable from benign nodes but lack genuine semantic contribution.
\end{itemize}
\textbf{Mechanism:} This module utilizes a ``Hardness-Gradient'' composite metric $\nabla_{\text{last}}L \cdot h_k$ to distinguish nodes semantically. Benign nodes exhibit a hardness $h_k \to 0$, whereas attackers maintain a high $h_k$ due to conflicts between the poisoning task and the main task.\\ 
\textbf{Output:} Based on the Shapley value $C_k^t$ (calculated via the algorithm in Section \ref{sec:chg_verification}), nodes with negative or negligible contribution values (associated with high hardness) are assigned zero weight, achieving precise mitigation of covert attacks.

\paragraph{Unified Trust Update \& Aggregation}
Regardless of the processing path, the final instantaneous contribution value ($C_k^t$) is integrated into the unified dynamic trust module.
\begin{itemize}
    \item \textbf{Long-term Reputation Modeling:} The system employs an Exponential Moving Average (EMA) to transform instantaneous contributions into a long-term reputation score $T_k^t$, thereby mitigating the impact of intermittent attacks.
    \item \textbf{Reputation Modulated Aggregation:} The final aggregation weight $w_{\text{final}}$ is derived from both the instantaneous contribution and the long-term reputation. The server applies this weight to $\Delta\theta_k^t$ to update the global model $\theta^{t+1}$.
\end{itemize}
This architecture ensures a smooth transition from morphological detection to contribution verification, guaranteeing robustness and fairness throughout the entire lifecycle of the Federated Learning process.

\section{Serial Statistical Filtering}
\label{sec:serial_filtering}

As discussed in Section \ref{sec:3.2}, the statistical filtering module within the proposed framework is strategically tailored for the \textit{Early Stage (Oscillation Period)} of Federated Learning (FL). During this phase, the primary objective is to establish a robust filtering mechanism to eliminate anomalous gradients that could otherwise derail the global optimization trajectory. While existing works offer various Byzantine-robust aggregation algorithms, such as Krum \cite{blanchard2017machine}, Median \cite{yin2018byzantine}, and RFA \cite{pillutla2022robust}, ensuring the systemic integrity of the Secure-CHG governance framework across the temporal spectrum—while balancing deployment feasibility with real-time monitoring—necessitates a streamlined approach. Consequently, we employ a serial filtering mechanism that integrates \textit{Cosine Similarity} with \textit{FL-Defender} as the preliminary defense module.

\subsection{Macro-Level Coarse Screening: Cosine Similarity Filtering}
\label{subsec:macro_screening}

This framework prioritizes \textit{Cosine Similarity} over \textit{Euclidean Distance} \cite{blanchard2017machine,yin2018byzantine,pillutla2022robust} as an initial defense to \textit{precisely intercept redirection attacks} (e.g., sign-flipping) that attempt to divert the global optimization trajectory.

\subsubsection{Constructing the Consensus Benchmark}
The server calculates the mean update vector $\Delta\theta^t = \frac{1}{K}\sum_{k=1}^{K}\Delta\theta_k^t$, representing the collective benign consensus direction for the current iteration.

\subsubsection{Directional Consistency Metric}
We evaluate the cosine similarity between each client's update $\Delta\theta_j^t$ and the consensus $\Delta\theta^t$:
\begin{equation}
\text{sim}(j,t) = \frac{\Delta\theta_j^t \cdot \Delta\theta^t}{\|\Delta\theta_j^t\| \cdot \|\Delta\theta^t\|}
\label{eq:cosine_sim}
\end{equation}

\subsubsection{Anomaly Determination}
Clients with a negative cosine similarity ($\text{sim}(j,t) < 0$) are excluded. A negative score indicates a directional deviation exceeding $90^\circ$, reliably signifying a malicious attempt to oppose the majority intent.

\subsection{Micro-Level Fine Screening: FL-Defender}
\label{subsec:micro_screening}

Macro screening may miss Targeted Poisoning Attacks (e.g., Backdoors) that embed malicious intent within a tiny fraction of dimensions to evade full-parameter statistical detection \cite{jebreel2023fl}. In high-dimensional models (e.g., ResNet-18), these minute alterations are easily masked by the vast majority of benign parameters, rendering full-model cosine similarity ineffective.

To address this, FL-Defender specifically analyzes \textit{Last-layer Gradients} via Principal Component Analysis (PCA), isolating micro-structural anomalies where attack intent is most concentrated.

\subsubsection{Theoretical Justification: Mathematical Necessity of Attack Intent}
For a $C$-class task, the predicted probability $p_i$ for class $i$ is:
\begin{equation}
p_i = \frac{e^{o_i}}{\sum_{j=1}^C e^{o_j}}
\end{equation}
Minimizing Cross-Entropy Loss $L(y, p)$ yields the last-layer weight gradient:
\begin{equation}
\nabla w_{i}^L = \frac{\partial L}{\partial w_{i}^L} = (p_i - y_i) \cdot a^{L-1}
\end{equation}
To misclassify samples, an attacker must alter the training target $y$, significantly deviating the error term $\delta_i = p_i - y_i$. This fundamental attack signature is strongest in the last layer and dilutes in earlier network layers.

\subsubsection{Noise Reduction via PCA}
To distinguish adversarial deviations from benign Non-IID data variations, FL-Defender applies PCA to the gradient similarity matrix. This provides two key benefits:
\begin{enumerate}
    \item \textbf{Suppression of Benign Variance:} Filters random high-dimensional perturbations as noise.
    \item \textbf{Amplification of Structural Deviation:} Highlights consistent malicious anomalies.
\end{enumerate}
Clients are then evaluated based on their deviation from the projected geometric centroid.

\subsubsection{Algorithm Implementation}
FL-Defender ensures a clean input distribution by filtering statistically anomalous structures during the early training phase. The procedure is detailed in Algorithm \ref{alg:fl_defender}.

\begin{algorithm}
\caption{FL-Defender Micro-Filtering}
\label{alg:fl_defender}
\begin{algorithmic}[1]
\REQUIRE Set of clients $\mathcal{K}$ (post macro-filtering), Last-layer gradients $\{\nabla_{\text{last}}L_j\}_{j \in \mathcal{K}}$, Previous trust history $H_{\text{prev}}$
\ENSURE Current trust $H_{\text{current}}$, Boolean flags $is\_benign$
\STATE \textbf{// Phase 1: Feature Engineering}
\FOR{$j \in \mathcal{K}$}
    \STATE $g_j = \text{L2\_Normalize}(\nabla_{\text{last}}L_j)$
\ENDFOR
\STATE Initialize Similarity Matrix $S \in \mathbb{R}^{|\mathcal{K}| \times |\mathcal{K}|}$
\FOR{$i=1$ to $|\mathcal{K}|$}
    \FOR{$j=1$ to $|\mathcal{K}|$}
        \STATE $S[i,j] = \text{CosineSimilarity}(g_i, g_j)$
    \ENDFOR
    \STATE $V_i = S[i,:]$ \COMMENT{Feature vector extraction}
\ENDFOR

\STATE \textbf{// Phase 2: Dimensionality Reduction}
\STATE $V_{all} = \text{Stack}(V_1, \dots, V_{|\mathcal{K}|})$
\STATE $V_{reduced} = \text{PCA}(V_{all}, \text{dims}=2)$ \COMMENT{Project to latent space}

\STATE \textbf{// Phase 3: Anomaly Scoring}
\STATE $centroid = \text{GeometricMedian}(V_{reduced})$
\FOR{$j \in \mathcal{K}$}
    \STATE $\zeta_j = \text{CosineSimilarity}(V_{reduced}[j], centroid)$
\ENDFOR

\STATE \textbf{// Phase 4: Trust Update \& Filtering}
\STATE Initialize $H_{\text{current}}$ as zeros
\FOR{$j \in \mathcal{K}$}
    \STATE $H_{\text{current}}[j] = \beta \cdot H_{\text{prev}}[j] + (1-\beta) \cdot \zeta_j$ \COMMENT{EMA smoothing}
\ENDFOR
\STATE $Q1 = \text{FirstQuartile}(H_{\text{current}})$
\FOR{$j \in \mathcal{K}$}
    \STATE $is\_benign[j] = (H_{\text{current}}[j] > Q1)$ \COMMENT{Threshold-based filtering}
\ENDFOR
\RETURN $H_{\text{current}}, is\_benign$
\end{algorithmic}
\end{algorithm}

FL-Defender robustly filters malicious clients using an Exponential Moving Average (EMA) mechanism to evaluate historical trust. By computing $H_{\text{current}}[j] = \beta \cdot H_{\text{prev}}[j] + (1-\beta) \cdot \zeta_j$, it acts as a low-pass filter to smooth benign Non-IID fluctuations while effectively preventing adaptive attackers from erasing their historical malicious footprint via intermittent camouflage.

While morphology-based statistical defenses (like FL-Defender) excel during the early stages of training, they inevitably fail upon model convergence. Therefore, we introduce an orthogonal, contribution-based mechanism—the CHG-Shapley algorithm (Section \ref{sec:chg_verification})—to guarantee full-lifecycle robustness.

\section{CHG Contribution Verification}
\label{sec:chg_verification}
In this section, we first provide a rigorous theoretical analysis of the ``Late-stage Failure'' inherent in existing morphology-based statistical defenses. Second, to overcome this critical bottleneck, we then propose an efficient contribution evaluation algorithm that explicitly amplifies the semantic differences between benign and malicious updates, restoring their separability even as gradient norms diminish.

\subsection{Theoretical Analysis of Late-stage Failure}
\label{sec:5.1}

Once the model is successfully purified and its convergence is accelerated during the early stages, the norms of benign gradients decay naturally. Capitalizing on this physical dynamic, adversaries immediately pivot into a stealth mode, thereby instigating a more insidious crisis defined as \textit{Late-stage Failure} (Definition \ref{def:1}).

As formalized in Definition \ref{def:1}, since $\lim_{t \to \infty} \|\Delta\theta_k^t\|_2 = 0$, the discrepancy $\delta_{\text{stat}}^t$ between malicious and benign updates in statistical features---such as geometric distance and directional consistency---approaches zero. Consequently, the detection sensitivity of any defense mechanism contingent on the morphology of $\Delta\theta$ plummets precipitously, ultimately culminating in complete failure. Fundamentally, attackers exploit the gradient decay properties inherent to the model's convergence process to achieve ``statistical invisibility.'' During this stage, a rational adversary will employ a ``covert fine-tuning'' strategy, deliberately compressing their update norms to a minuscule magnitude of the same order as benign updates (i.e., $\epsilon$-bounded). 

This renders conventional measurement metrics entirely ineffective. In the following, we elucidate this multi-faceted failure across three dimensions: Euclidean distance, directional metrics, and microscopic discrimination:

\textbf{Failure of Euclidean Distance-Based Defenses (e.g., Krum)} 

Consider a $d$-dimensional parameter space. Algorithms like Krum rely on computing the Euclidean distance $\|u-v\|_2$ between update vectors. In the late convergence stage, the norm of benign updates decays such that $\|\Delta\theta_{\text{benign}}\|_2 \approx \sqrt{d}\sigma_t \to 0$. The expected squared distance between benign nodes $i$ and $j$ is $\mathbb{E}[\|\Delta\theta_i - \Delta\theta_j\|_2^2] = 2d\sigma_t^2$. To evade detection, a malicious node $m$ constrains its update within the neighborhood of benign updates, establishing $\|\delta\| \approx \sqrt{d}\sigma_t$. Consequently, the expected distance between a malicious node and a benign node, $\mathbb{E}[\|\Delta\theta_i - \Delta\theta_m\|_2^2]$, falls into the exact same order of magnitude $\mathcal{O}(d\sigma_t^2)$ as the intra-class distance among benign nodes. Governed by the concentration of measure phenomenon in high-dimensional spaces, when $\sigma_t$ approaches infinitesimal levels, the overlap region between the benign and malicious distributions is maximized. This homogenization completely prevents distance-ranking defense algorithms from distinguishing adversaries from benign statistical noise.

\textbf{Failure of Directional Metrics (Cosine Similarity)}

Cosine similarity is defined as $\text{sim}(u,v) = \frac{u \cdot v}{\|u\|\|v\|}$. In the late convergence stage, the global average update $\Delta\theta_{\text{avg}} \approx \frac{1}{|\mathcal{B}|}\sum \Delta\theta_{\text{benign}}$ degenerates toward random noise tightly clustered around the zero vector. As $\|\Delta\theta_{\text{avg}}\| \to 0$, cosine similarity becomes extremely sensitive to minuscule perturbations. If the adversary generates a vector $\delta$ that perfectly mimics zero-mean random noise, then: $\lim_{t \to \infty} P(\text{sim}(\Delta\theta_{\text{mal}}, \Delta\theta_{\text{avg}}) < 0) \approx P(\text{sim}(\Delta\theta_{\text{benign}}, \Delta\theta_{\text{avg}}) < 0) \approx 0.5$. At this critical juncture, the cosine similarity distributions of malicious and benign nodes perfectly overlap, rendering simple sign thresholds or numerical truncation indistinguishable. This signifies that the macroscopic rough screening discussed in Section \ref{subsec:macro_screening} fundamentally loses its requisite statistical baseline.

\textbf{Failure of Microscopic Discrimination}

Concurrently, the FL-Defender microscopic fine screening detailed in Section \ref{subsec:micro_screening} encounters an identical detection bottleneck. In the late convergence stage, the feature extraction layers stabilize, yielding an exceedingly low variance for benign gradients. Here, attackers only need to inject vanishingly weak trigger signals to sustain the backdoor. These subtle signals conceal themselves within a massive parameter background, no longer manifesting as pronounced principal component anomalies. The ``structural deviations'' that FL-Defender attempts to extract via Principal Component Analysis (PCA) are effortlessly submerged by the inherent heterogeneous noise of Non-IID data distributions. 

Formally, assume the distribution of benign gradients is $\Delta\theta_{\text{benign}} \sim \mathcal{N}(0, \sigma_t^2 I)$, where $\sigma_t$ converges to an infinitesimal value $\epsilon$. Covert attackers construct their updates as $\Delta\theta_{\text{mal}} = \delta$, subject to $\|\delta\| \approx \epsilon$. Let $\Sigma$ denote the covariance matrix of the last-layer gradients. In the early stages, the attack signal $S$ is dominant such that $\|\Sigma_{\text{attack}}\| \gg \|\Sigma_{\text{noise}}\|$, allowing the attack features to dictate the primary principal component $v_1$. However, in the late stage, the benign gradient variance $\sigma_t^2$ collapses, and attackers strategically compress their signal strength to $\|\delta\| \le \sigma_t$ for stealth. Consequently, the principal components extracted by PCA are predominantly dictated by the intrinsic noise variance of Non-IID data rather than adversarial features. The projection $P(\Delta\theta_{\text{mal}})$ of malicious samples in the principal component space strictly satisfies:
\begin{equation}
\|P(\Delta\theta_{\text{mal}}) - \text{Centroid}\|_2 \le \|P(\Delta\theta_{\text{benign}}) - \text{Centroid}\|_2
\end{equation}
This inequality proves that malicious samples fall squarely within the confidence interval of benign nodes, rendering FL-Defender's anomaly detection mathematically obsolete.

In summary, the essence of Late-stage Failure lies in the morphological homogenization of updates: as the global model converges and gradient norms vanish, stealthy malicious updates become statistically indistinguishable from benign ones. This fundamental collapse of geometric defenses directly triggers a catastrophic surge in the Attack Success Rate (ASR) of advanced poisoning techniques.

\subsection{CHG-Shapley: Efficient Contribution Evaluation Algorithm}

As discussed in Section \ref{sec:5.1}, the fundamental cause of Late-stage Failure is the collapse of geometric separability as gradient norms diminish. An intuitive solution to resolve this is to explicitly amplify the direct semantic differences. We observe that malicious nodes inherently struggle to fit their anomalous poisoned samples, causing their associated training loss---or Hardness ($h_i$)---to remain persistently higher than that of regular nodes. Furthermore, their gradient update directions inevitably diverge from the benign consensus. Therefore, Secure-CHG shifts the defense paradigm from analyzing superficial ``Statistical Morphology'' to verifying genuine contribution values based on this intrinsic semantic difference.

By strategically coupling the persistent hardness with the gradient, Secure-CHG projects the updates into a novel ``Hardness-Gradient'' Semantic Space. In this space, the scalar hardness acts as a natural amplifier for malicious intent. Even as the raw geometric distance between parameter updates collapses ($\lim_{t \to \infty} \|\Delta\theta_i - \Delta\theta_m\|_2 = 0$), this composite mechanism successfully magnifies the subtle adversarial traces, decisively highlighting and isolating the malicious nodes from the benign cluster. This restorative separability is visually and empirically corroborated by the distinct clustering transitions shown in Fig. \ref{fig:tsne}(b) and \ref{fig:tsne}(c).

Ultimately, by explicitly incorporating the semantic variable $h_i$, CHG-Shapley maintains a non-zero statistical distance between benign actors and malicious actors, effectively circumventing the theoretical limitations of geometric defenses in the late training stage.

To translate this restored separability within the ``Hardness-Gradient'' semantic space into practically quantifiable node contributions, the crux lies in seamlessly embedding it into the Shapley value evaluation framework. To achieve this, we must establish a utility function $U(S)$ that accurately reflects the authentic value of a data subset $S$. Traditionally, $U(S)$ is defined as the model's accuracy on a validation set after retraining on $S$, which is computationally prohibitive in federated settings.

To circumvent the need for retraining, a common alternative is to evaluate the model's training loss reduction caused by the parameter update. Let the global loss be $f(\theta) := \frac{1}{N_{\text{total}}}\sum_{i \in N} f(i;\theta)$. Consider an update step using a generic aggregated raw gradient $g = \frac{1}{|S|}\sum_{i \in S} \nabla f(i;\theta)$, yielding new parameters $\theta_g = \theta - \eta g$.

\textbf{Lemma 1 (Nesterov, 2018):} For a differentiable function $f$ with $M$-Lipschitz continuous gradients, if $\eta = 1/M$, the loss reduction is bounded by:
\begin{equation}
f(\theta_g) \le f(\theta) - \frac{1}{2M} \|\nabla f(\theta)\|_2^2 + \frac{1}{2M} \|\nabla f(\theta) - g\|_2^2
\label{eq:nesterov_bound}
\end{equation}

Equation (\ref{eq:nesterov_bound}) implies that the loss reduction is fundamentally tied to the alignment between the update direction $g$ and the global gradient $\nabla f(\theta)$. This mathematical property naturally suggests a retraining-free proxy utility: $C(S) \propto \|\nabla f(\theta)\|_2^2 - \|\nabla f(\theta) - g\|_2^2$.

However, here lies a bottleneck: relying on raw gradients $g$ falls victim to the Late-stage Failure. As the model converges, raw gradients vanish, and the proxy utility $C(S)$ completely loses its discriminative power against stealthy adversaries.

To break this bottleneck and evaluate genuine contributions post-convergence, we must project this proxy framework into the ``Hardness-Gradient'' space. Instead of using raw vanishing gradients, we introduce sample hardness $h_i = f(i;\theta)$ to explicitly scale the local gradients, defining a composite vector $x_i = h_i \nabla f(i;\theta)$. Because malicious samples inherently maintain persistently high hardness, this semantic scaling amplifies adversarial traces and prevents their vectors from vanishing. By replacing the raw gradients with the hardness-weighted vectors (i.e., mapping $g \to \frac{1}{|S|}\sum_{i \in S} x_i$ and $\nabla f(\theta) \to \alpha = \frac{1}{N}\sum_{i \in N} x_i$), we construct our final utility function, the \textbf{CHG Score}:
\begin{equation}
U(S) = \|\alpha\|_2^2 - \left\| \frac{1}{|S|} \sum_{i \in S} x_i - \alpha \right\|_2^2
\label{eq:chg_score_simplified}
\end{equation}
where $x_i = h_i \nabla f(i;\theta)$ and $\alpha = \frac{1}{N}\sum_{i \in N} x_i$. This utility function rewards subsets whose update directions align with the global optimization direction weighted by sample hardness.

To bridge this theoretical formulation with practical Federated Learning deployments, we must circumvent the prohibitive communication overhead of transmitting instance-level, full-parameter gradients. Therefore, we explicitly operationalize this metric at the client level. Specifically, we define the direct input to the CHG-Shapley algorithm as the composite ``Hardness-Gradient'' vector $x_k = \nabla_{\text{last}}L_k \cdot h_k$. By feeding this layer-specific, hardness-scaled semantic vector into the evaluation module as the definitive input, we effectively compress the required state space while fully preserving the distinct adversarial traces. Consequently, this utility function elegantly rewards clients whose structural update directions align with the global trajectory, heavily scaled by their localized sample hardness.

Based on the derivation in Appendix A, we derive a Closed-Form Solution for the Shapley value based on the proposed quadratic utility function. Leveraging the linearity axiom, we decompose $U(S)$ into linear combinations of basic functions and derive the marginal contributions analytically.

\textbf{Theorem 1:} Assuming the utility function $U(S)$ defined in Eq. (\ref{eq:chg_score_simplified}), the Shapley value $\phi_k(U)$ for the $k$-th client is given by the following closed-form expression:

\begin{figure*}[t]
\begin{equation}
\label{eq:closed_form}
\resizebox{0.95\hsize}{!}{$
\begin{aligned}
\phi_k(U) = \sum_{j \in D_k} \Bigg[ 
& \left( -\frac{1}{n_k}\sum_{l=1}^{n_k} \frac{1}{l^2} + \frac{1}{n_k(n_k-1)}\left(2\sum_{l=1}^{n_k}\frac{1}{l} - 3\sum_{l=1}^{n_k}\frac{1}{l^2} + \frac{1}{n_k}\right) + \frac{2\left(2\sum_{l=1}^{n_k}\frac{1}{l} - 2\sum_{l=1}^{n_k}\frac{1}{l^2} - 1 + \frac{1}{n_k}\right)}{n_k(n_k-1)(n_k-2)} \right) \|x_j\|_2^2 \\
& - \frac{2\left(\sum_{l=1}^{n_k}\frac{1}{l} - \sum_{l=1}^{n_k}\frac{1}{l^2} - \frac{1}{n_k} + \frac{1}{n_k^2}\right)}{(n_k-1)(n_k-2)} \langle \sum_{i \in N} x_i, x_j \rangle \\
& + \frac{2\sum_{l=1}^{n_k}\frac{1}{l} - 2\sum_{l=1}^{n_k}\frac{1}{l^2} - 1 + \frac{1}{n_k}}{n_k(n_k-1)(n_k-2)} \left\|\sum_{i \in N} x_i\right\|_2^2 \\
& + \left( \frac{\sum_{l=1}^{n_k}\frac{1}{l^2} - \frac{1}{n_k}}{n_k(n_k-1)} - \frac{2\sum_{l=2}^{n_k}\frac{1}{l} - 2\sum_{l=2}^{n_k}\frac{1}{l^2} - 1 + \frac{1}{n_k}}{n_k(n_k-1)(n_k-2)} \right) \left( \sum_{i \in N} \|x_i\|_2^2 \right) \\
& + \frac{2(\sum_{l=1}^{n_k}\frac{1}{l} - \frac{1}{n_k})}{n_k-1} \langle x_j, \alpha \rangle - \frac{2(\sum_{l=1}^{n_k}\frac{1}{l} - 1)}{n_k(n_k-1)} \left\langle \sum_{i \in N} x_i, \alpha \right\rangle
\Bigg]
\end{aligned}
$}
\end{equation}
\end{figure*}

\begin{remark}
Eq. (\ref{eq:closed_form}) represents the closed-form structural solution derived from combinatorial expansion, which involves harmonic sums over the dataset size $n_k$. This bypasses the exponential complexity of traditional Shapley computation.

Operationally, CHG-Shapley functions as a single-round marginal contribution proxy evaluated along the realized training path. It efficiently captures how a client's update enhances the model's capacity to generalize on difficult samples without requiring exhaustive coalition enumeration.
\end{remark}

\section{Unified Trust Update and Aggregation}
\label{sec:unified_trust}

Building on the CHG-Shapley contribution scores $C_k^t$ derived in Section \ref{sec:chg_verification}, this section introduces a dynamic reputation assessment mechanism to precisely identify malicious clients. Ultimately, these reputation scores are integrated into the final weighted aggregation to completely neutralize adversarial impacts.

\subsection{Contribution-based Dynamic Reputation Assessment}
\label{subsec:dynamic_reputation}

\subsubsection{Logarithmic Scaling}
To mitigate the influence of outliers and compress the dynamic range of contribution values, we apply a logarithmic scaling function to the instantaneous contributions $C_k^t$ of benign clients that have successfully passed the defense filtering (as detailed in Section \ref{sec:serial_filtering}):
\begin{equation}
w_{raw}^t(k) = \log(1 + \max(C_k^t, \epsilon))
\label{eq:log_scaling}
\end{equation}
Here, $\epsilon$ is a small constant (e.g., $10^{-9}$) introduced to ensure numerical stability.

\subsubsection{Relative Contribution Filtering}
To further marginalize low-quality nodes that evade initial defense mechanisms but offer negligible contributions, we implement a relative performance-based filter.
First, a baseline adjustment is performed by subtracting a scaled minimum contribution:
\begin{equation}
w_{adj}^t(k) = \max\left(0, w_{raw}^t(k) - \gamma \cdot \min_{i}(w_{raw}^t(i))\right)
\end{equation}
where $\gamma$ is a scaling factor close to 1 (e.g., 0.99).
Second, we compute the mean of the adjusted weights, $\bar{w}_{adj}^t$, and establish a quality threshold. Any client with $w_{adj}^t(k)$ falling below this threshold (e.g., $0.5 \cdot \bar{w}_{adj}^t$) is classified as a low-quality contributor for the current round, and their weight is reset to zero.
Following these filtering steps, the remaining weights are normalized to derive the instantaneous normalized weights, $w_{inst}^t(k)$, for the current round.

\subsubsection{Long-term Reputation (Trust) Update}
The core of the dynamic reputation mechanism relies on a long-term reputation score, $T_k^t$, maintained for each client $k$. This score is updated using an Exponential Moving Average (EMA) to smooth short-term fluctuations and build trust based on historical behavior:
\begin{equation}
T_k^t = \beta T_k^{t-1} + (1-\beta) w_{inst}^t(k)
\label{eq:trust_update}
\end{equation}
where $\beta \in [0, 1]$ is the decay factor. A higher $\beta$ places greater emphasis on historical performance, thereby stabilizing the reputation score against transient anomalies.

\subsubsection{Trust-Modulated Aggregation Weight}
The final aggregation weight, $w_{trust}^t(k)$, for client $k$ in round $t$ is determined by a linear combination of its instantaneous contribution, $w_{inst}^t(k)$, and its deviation from the population's average reputation, $T^t$. Specifically:
\begin{equation}
w_{trust}^t(k) = w_{inst}^t(k) + \lambda(T_k^t - T^t)
\label{eq:trust_modulation}
\end{equation}
where $\lambda$ represents the reputation modulation rate. This formulation ensures that a client with a superior historical reputation ($T_k^t > T^t$) receives a weight incentive, effectively buffering against minor fluctuations in instantaneous contribution. Conversely, clients with poor historical performance are penalized.

By enforcing non-negativity via clipping ($w_{final}^t(k) = \max(0, w_{trust}^t(k))$) and performing a final normalization, we obtain the definitive weights for the global model update. This mechanism seamlessly integrates defense (filtering $C_k^t=0$), evaluation (computing $C_k^t$), and incentives (deriving $w_{final}^t(k)$), thereby establishing a robust and adaptive federated governance loop.

\subsection{Reputation Modulated Aggregation Protocol}

Building upon the modulated weights $w_{trust}^t(k)$, this section describes the final secure aggregation protocol. The objective is to translate abstract reputation scores into concrete model updates, ensuring the global model assimilates knowledge exclusively from high-reputation, high-contribution nodes.

\subsubsection{Normalization and Global Model Update}
To constrain the aggregated model parameters within a valid numerical range and prevent divergence due to weight explosion, we apply a Rectified Linear Unit (ReLU) operation followed by normalization to derive the final aggregation coefficients $w_{final}^t(k)$:
\begin{equation}
w_{final}^t(k) = \frac{\max(0, w_{trust}^t(k))}{\sum_{j \in \mathcal{K}} \max(0, w_{trust}^t(j))}
\label{eq:final_normalization}
\end{equation}
The denominator represents the sum of modulated weights across all clients. In the extreme case where the denominator is zero (indicating all clients are deemed malicious or non-contributory), the update is skipped to preserve model integrity.

Subsequently, the server utilizes these final weights to compute the weighted average of the update vectors $\Delta\theta_k^t$, resulting in the global model update:
\begin{equation}
\theta^{t+1} \leftarrow \theta^t + \eta_g \sum_{k=1}^K w_{final}^t(k) \cdot \Delta\theta_k^t
\label{eq:global_update}
\end{equation}
where $\eta_g$ denotes the global learning rate. This aggregation mechanism provides dual security guarantees:
\begin{itemize}
    \item \textbf{Malicious Exclusion:} Nodes identified by defense mechanisms or exhibiting negligible contributions are assigned $w_{final}^t(k) \approx 0$, effectively blocking the injection of malicious gradients.
    \item \textbf{Benign Incentive:} Benign nodes with consistent historical performance and significant current contributions receive amplified weights, thereby accelerating convergence toward the optimal solution.
\end{itemize}

\subsubsection{Algorithm Description}
Overall, the unified aggregation process is summarized in Algorithm \ref{alg:unified_aggregation}.

\begin{algorithm}
\caption{Secure-CHG Unified Trust Aggregation}
\label{alg:unified_aggregation}
\begin{algorithmic}[1]
\REQUIRE Client updates $\{\Delta\theta_k^t, \nabla_{\text{last}}L_k, h_k\}_{k \in \mathcal{K}}$; Previous global model $\theta^t$; Previous trust scores $T^{t-1}$; Current stage flag $State$ (Early/Late).
\ENSURE New global model $\theta^{t+1}$; Updated trust scores $T^t$.
\STATE Initialize: $w_{inst}^t \leftarrow \text{zeros}(K)$, $Mask^t \leftarrow \text{ones}(K)$
\STATE \textbf{// Phase 1: Dual-Mode Defense \& Evaluation}
\FOR{each client $k \in \mathcal{K}$}
    \IF{$State$ is Early Stage}
        \IF{$\text{Cosine}(\Delta\theta_k^t, \Delta\theta_{\text{avg}}) < 0$ \textbf{or} $\text{FL-Defender}(\nabla_{\text{last}}L_k) == \text{Malicious}$}
            \STATE $Mask^t[k] \leftarrow 0$; \textbf{continue} \COMMENT{Statistical Filtering}
        \ENDIF
    \ENDIF
    \STATE Calculate $C_k^t \leftarrow \text{CHG-Shapley}(\Delta\theta_k^t, h_k)$ using Theorem 1
\ENDFOR
\STATE \textbf{// Phase 2: Dynamic Trust Assessment}
\FOR{each client $k \in \mathcal{K}$ where $Mask^t[k]=1$}
    \STATE $w_{raw}^t(k) \leftarrow \log(1 + \max(C_k^t, \epsilon))$ \COMMENT{Log Scaling}
    \STATE $w_{adj}^t(k) \leftarrow \max(0, w_{raw}^t(k) - \gamma \cdot \min(w_{raw}^t))$ \COMMENT{Relative Filtering}
\ENDFOR
\STATE Normalize $w_{adj}^t$ to get $w_{inst}^t$
\STATE Update Trust: $T_k^t \leftarrow \beta T_k^{t-1} + (1-\beta) w_{inst}^t(k)$ \COMMENT{Eq. (\ref{eq:trust_update})}
\STATE \textbf{// Phase 3: Reputation Modulated Aggregation}
\STATE Calculate Average Trust $T_{avg}^t \leftarrow \text{Mean}(T^t)$
\FOR{each client $k \in \mathcal{K}$}
    \STATE $w_{trust}^t(k) \leftarrow w_{inst}^t(k) + \lambda(T_k^t - T_{avg}^t)$ \COMMENT{Eq. (\ref{eq:trust_modulation})}
\ENDFOR
\STATE Normalize $w_{trust}^t$ to get final weights $w_{final}^t$ \COMMENT{Eq. (\ref{eq:final_normalization})}
\STATE Update Global Model: $\theta^{t+1} \leftarrow \theta^t + \eta_g \sum w_{final}^t(k) \Delta\theta_k^t$ \COMMENT{Eq. (\ref{eq:global_update})}
\RETURN $\theta^{t+1}, T^t$
\end{algorithmic}
\end{algorithm} 

\subsection{System Stability Analysis}
\label{subsec:stability}

While the dynamic weight adjustment mechanism enhances fairness and security, it introduces the potential for weight fluctuations that could destabilize training. To guarantee convergence stability, we analyze the Secure-CHG framework across three dimensions: the smoothness of trust updates, the boundedness of aggregation weights, and the hysteresis of defense state switching.

\subsubsection{Smoothness of Trust Updates}
In Federated Learning, the stochastic nature of mini-batch sampling causes inevitable statistical fluctuations in a client's single-round contribution $w_{inst}^t$. Directly using instantaneous contributions as weights would induce severe oscillations in the optimization trajectory.
Secure-CHG mitigates this via the Exponential Moving Average (EMA) mechanism (Eq. (\ref{eq:trust_update})), which functions as a low-pass filter. For any time $t$, the variation in trust value is constrained by the decay factor $\beta$:
\begin{equation}
|T_k^t - T_k^{t-1}| = (1-\beta) |w_{inst}^t(k) - T_k^{t-1}|
\end{equation}
By setting $\beta$ close to 1 (e.g., 0.9), the system suppresses single-round anomalies---whether from random noise or adversarial attempts. This imparts ``reputation inertia,'' preventing the weights of benign nodes from collapsing due to transient low-quality updates while simultaneously preventing attackers from rapidly accruing trust through sporadic high-quality disguises.

\subsubsection{Boundedness of Aggregation Weights}
Stability is further ensured by the boundedness of the aggregation weights. In Eq. (\ref{eq:trust_modulation}), the trust modulation term is $\lambda(T_k^t - T^t)$. Since $w_{inst}$ is normalized ($\in [0,1]$), the long-term reputation $T_k^t$ is similarly constrained to the $[0,1]$ interval.
Consequently, the deviation of the modulated weight $w_{trust}^t$ is bounded within $[-\lambda, 1+\lambda]$. When combined with the ReLU non-negative clipping and final normalization in Eq. (\ref{eq:final_normalization}), we ensure:
\begin{equation}
\sum_{k=1}^K w_{final}^t(k) = 1 \quad \text{and} \quad w_{final}^t(k) \ge 0
\end{equation}
This satisfies the convexity condition, guaranteeing that the global model update $\Delta\theta^{t+1}$ remains within the convex hull of the local update vectors. Theoretically, this prevents the aggregation process from causing the model parameters to diverge into an invalid space.

\subsubsection{Hysteresis in Mode Switching}
Beyond numerical stability, strategic stability is paramount. As described in Section \ref{sec:3.2}, the framework incorporates a hysteresis mechanism for switching between ``Statistical Filtering'' and ``Contribution Verification'' modes. This design prevents the system from oscillating (chattering) between distinct defense logics due to minor loss fluctuations near the convergence threshold. Such strategic stability ensures the continuity of the optimization trajectory, facilitating a smooth transition into the convergence phase.

\section{Experiments}
\label{sec:experiments}

This section presents systematic experiments to verify the effectiveness of the Secure-CHG framework against strategic poisoning attacks. The core objectives are: 1) to verify the ``Late-stage Failure'' defect of traditional defense mechanisms; 2) to verify the ``Early-stage Vulnerability'' of pure contribution evaluation mechanisms; and 3) to demonstrate the robustness of the proposed hybrid framework (Secure-CHG) throughout the entire training lifecycle.

\subsection{Experimental Setup and Baselines}
Experiments were conducted on the CIFAR-10 dataset using the LeNet model, simulating a Federated Learning scenario with 10 clients (including 2 malicious clients performing ``dog $\to$ cat'' label-flipping attacks). The EMA decay factor $\beta$ in FL-Defender was set to 0.8, and $\beta$ for long-term reputation updates was set to 0.9.

We established comprehensive baselines, including:
\begin{itemize}
    \item \textbf{No Defense:} FedAvg.
    \item \textbf{Traditional Statistical Defenses:} FedMedian, TrimmedMean, MultiKrum, FLTrust.
    \item \textbf{Ablation Baselines:} Pure contribution evaluation (Only\_CHG) and statistical filtering only (Only\_series\_filters).
\end{itemize}

The core evaluation metrics, all computed as averages over the final 30 rounds, are:
\begin{itemize}
    \item \textbf{Average Global Model Accuracy (Avg.\ All-Acc):} Measures the final generalization performance.
    \item \textbf{Average Attack Success Rate (Avg.\ ASR):} The primary indicator of defense robustness (lower is better).
    \item \textbf{Average Source Category Accuracy (Avg.\ Src-Acc):} The proportion of correctly classified samples from attacked source categories.
\end{itemize}

\subsection{Results and Analysis}

This section details the analysis of the experimental data to systematically verify the core arguments of this paper. All performance metrics are average values from the late training stage (Rounds 21-50) to evaluate final robustness and performance during the convergence phase.

\begin{table}[htbp]
    \centering
    \caption{Performance Comparison of Different Defense Mechanisms (Late Stage Avg.)}
    \label{tab:performance_comparison}
    \resizebox{\columnwidth}{!}{%
    \begin{tabular}{lccc}
        \toprule
        \textbf{Method} & \textbf{Avg.\ All-Acc (\%)} & \textbf{Avg.\ Src-Acc (\%)} & \textbf{Avg.\ ASR (\%)} \\
        \midrule
        FedAvg & 60.45 & 19.52 & 47.12 \\
        FedMedian & 60.45 & 28.03 & 34.82 \\
        TrimmedMean & 62.35 & 31.28 & 37.80 \\
        MultiKrum & 61.14 & 27.56 & 41.97 \\
        FLTrust & 62.40 & 19.91 & 50.84 \\
        Only\_CHG & 52.62 & 0.01 & 67.31 \\
        Only\_series\_filters & 61.29 & 35.79 & 33.06 \\
        \textbf{Filter\_CHG (Secure-CHG, ours)} & \textbf{60.50} & \textbf{49.87} & \textbf{18.84} \\
        Krum\_CHG & 61.25 & 48.20 & 20.82 \\
        \bottomrule
    \end{tabular}%
    }
\end{table}

\begin{figure}[htbp]
    \centering
    \includegraphics[width=\linewidth]{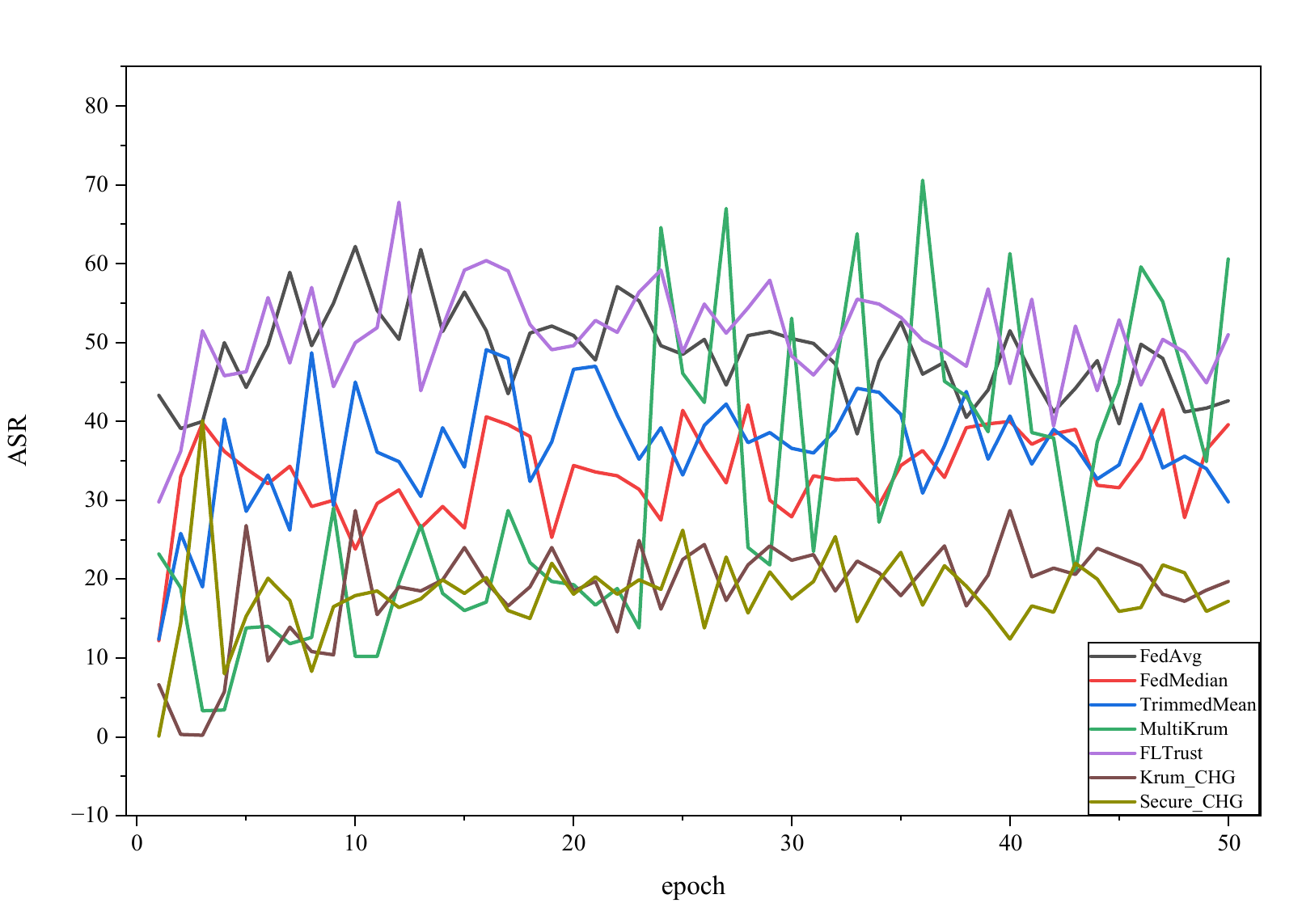}
    \caption{Trend of Attack Success Rate (ASR) over training epochs.}
    \label{fig:asr_trend}
\end{figure}

\begin{figure}[htbp]
    \centering
    \includegraphics[width=\linewidth]{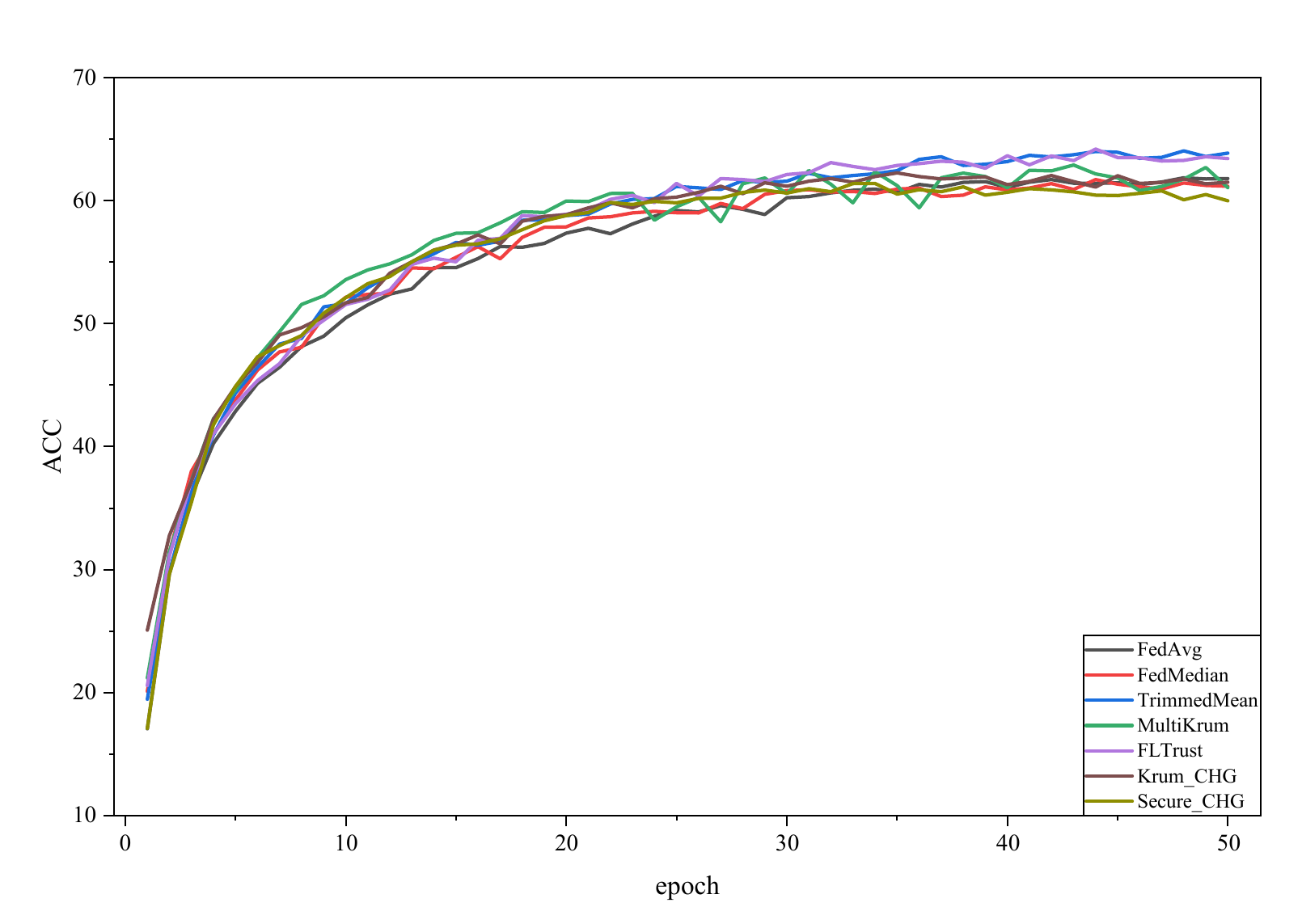}
    \caption{Trend of Main Task Accuracy (ACC) over training epochs.}
    \label{fig:acc_trend}
\end{figure}

\begin{figure}[htbp]
    \centering
    \includegraphics[width=\linewidth]{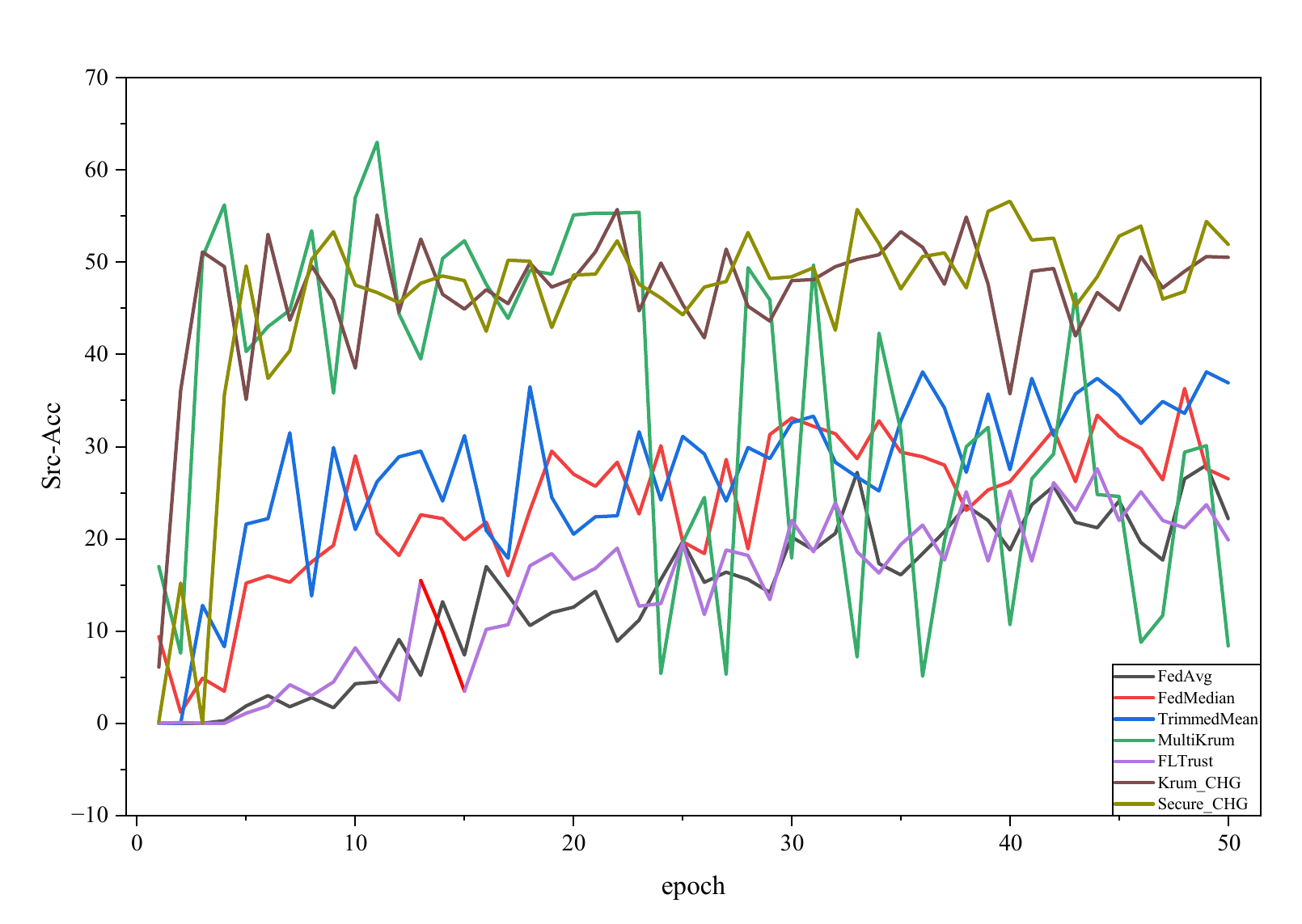}
    \caption{Trend of Source Class Accuracy (Src-Acc) over training epochs.}
    \label{fig:src_acc_trend}
\end{figure}

First, traditional statistical defenses universally suffer from \textbf{``Late-stage Failure.''} As shown in Fig.~\ref{fig:asr_trend} and Table~\ref{tab:performance_comparison}, in the late training stages, the ASR curves of methods like MultiKrum exhibit significant fluctuations or even rebounds. Its final Avg.\ ASR (41.97\%) shows limited improvement compared to the undefended baseline A\_FedAvg (47.12\%). This confirms that when the model converges and update vectors tend to be identical, defense mechanisms relying on statistical morphology tend to fail.

\begin{figure*}[htbp]
    \centering
    \subfloat[Early stage separability]{\includegraphics[width=0.32\linewidth]{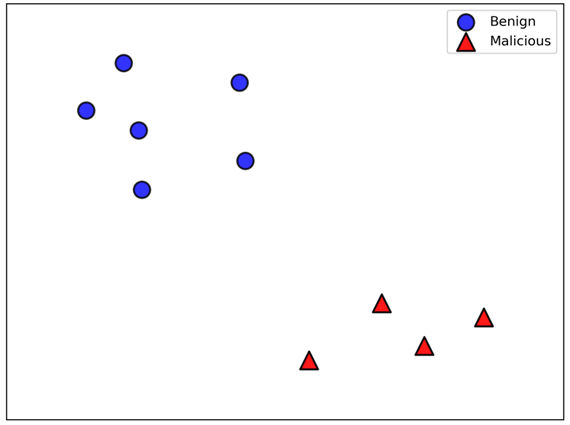}}\hfill
    \subfloat[Late stage collapse under raw gradients]{\includegraphics[width=0.32\linewidth]{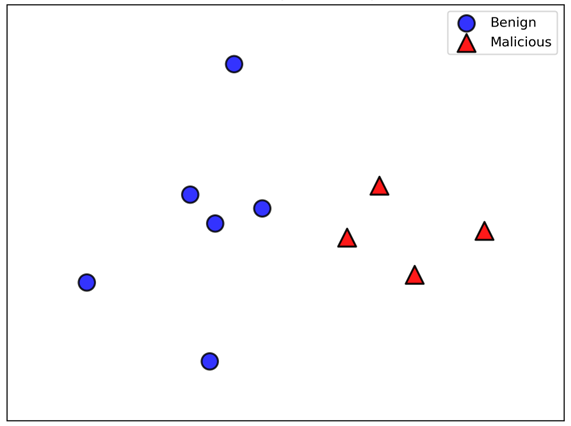}}\hfill
    \subfloat[Late stage separability restoration under Secure-CHG space]{\includegraphics[width=0.32\linewidth]{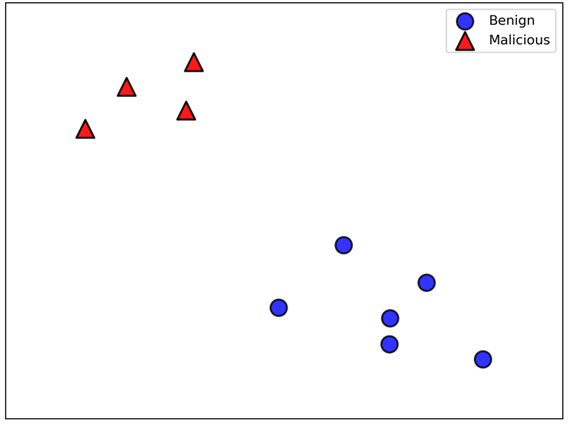}}
    \caption{Visualization of feature space evolution using t-SNE.}
    \label{fig:tsne}
\end{figure*}

To visually verify the gradient evolution laws described in \textit{Definition 1} (Section~\ref{3.1}) and demonstrate the discriminative ability of Secure-CHG in feature space, we utilized the t-SNE technique to map high-dimensional model update vectors to a 2D plane. Fig.~\ref{fig:tsne} displays the distribution of benign clients (blue circles) and malicious clients (red triangles) at different training stages.

As shown in Fig.~\ref{fig:tsne}, the visual evolution of the feature space intuitively reveals the efficacy boundaries of defense mechanisms. In the early stage (Round 5, Fig.~\ref{fig:tsne}(a)), malicious updates manifest as significant \textit{Statistical Outliers} due to drastic perturbations, creating a clear linear decision boundary with benign nodes. However, as the model enters the convergence period (Round 45, Fig.~\ref{fig:tsne}(b)), benign gradients decay toward random noise, and attackers exploit this to simulate mainstream directions. This leads to a \textit{High Overlap} where malicious nodes penetrate benign clusters, proving that defenses relying purely on geometric morphology fail completely at this stage.

Addressing this dilemma, Fig.~\ref{fig:tsne}(c) shows feature reconstruction after introducing Secure-CHG. Despite morphological convergence, malicious samples exhibit higher hardness (Loss) due to the conflict between the ``poisoning task'' and the ``main task.'' This \textit{Semantic Discrepancy}, amplified by the CHG mapping ($h_k \nabla f(\theta)$), effectively isolates malicious nodes from benign clusters. This strongly proves that Secure-CHG can break through geometric camouflage limits and restore linear separability in the late convergence stage.

\subsection{Ablation Studies}
To explicitly elucidate the individual contribution of each module, we constructed ablation studies featuring two variants: Only\_CHG (which solely relies on CHG-Shapley for contribution evaluation prior to aggregation) and Only\_series\_filters (which exclusively employs FL-Defender for filtering, followed by naive FedAvg aggregation).

\begin{figure}[H]
    \centering
    \includegraphics[width=\linewidth]{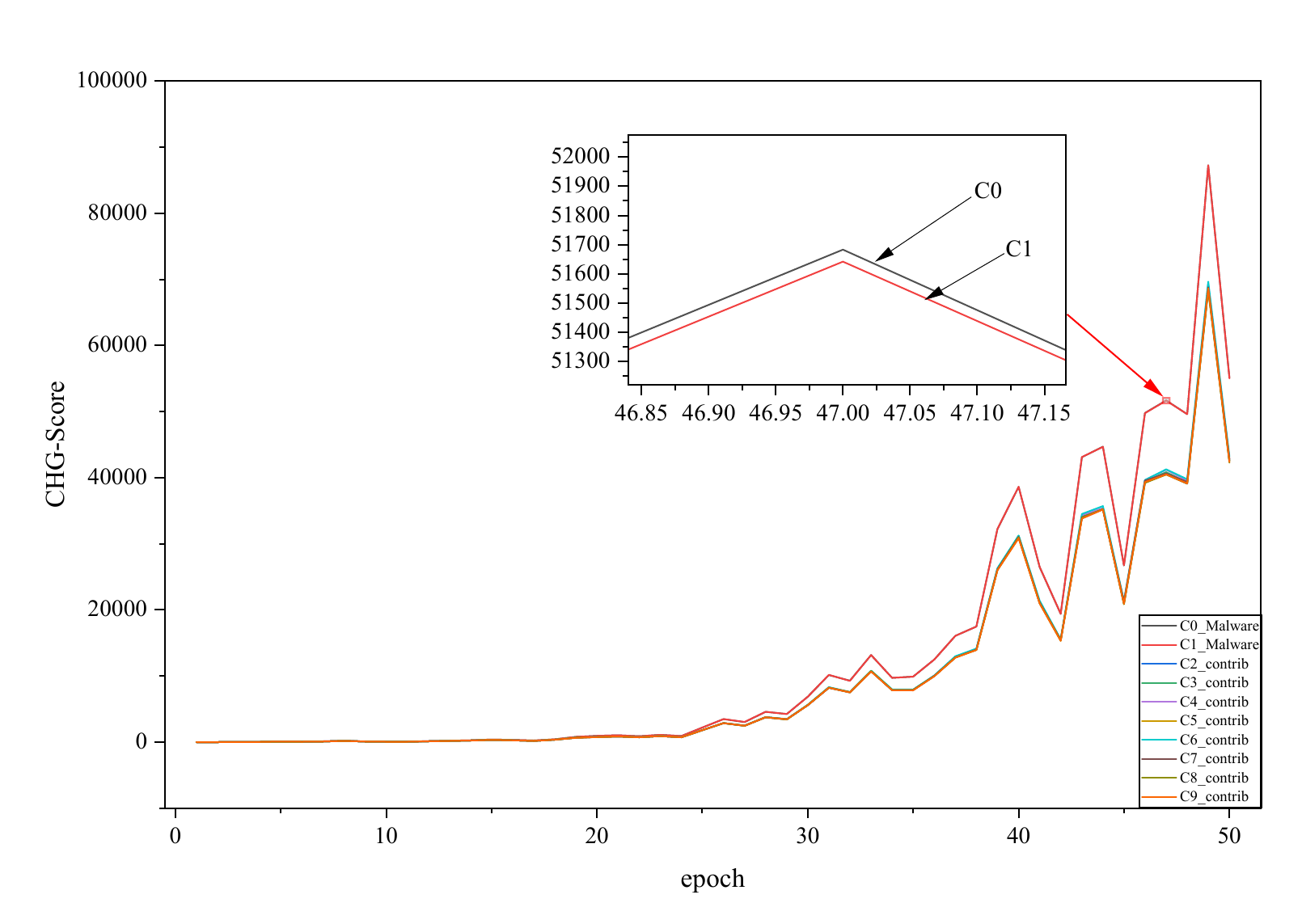}
    \caption{Trend of CHG contribution values over time for Only\_CHG.}
    \label{fig:chg_only_trend}
\end{figure}

\begin{figure}[H]
    \centering
    \includegraphics[width=\linewidth]{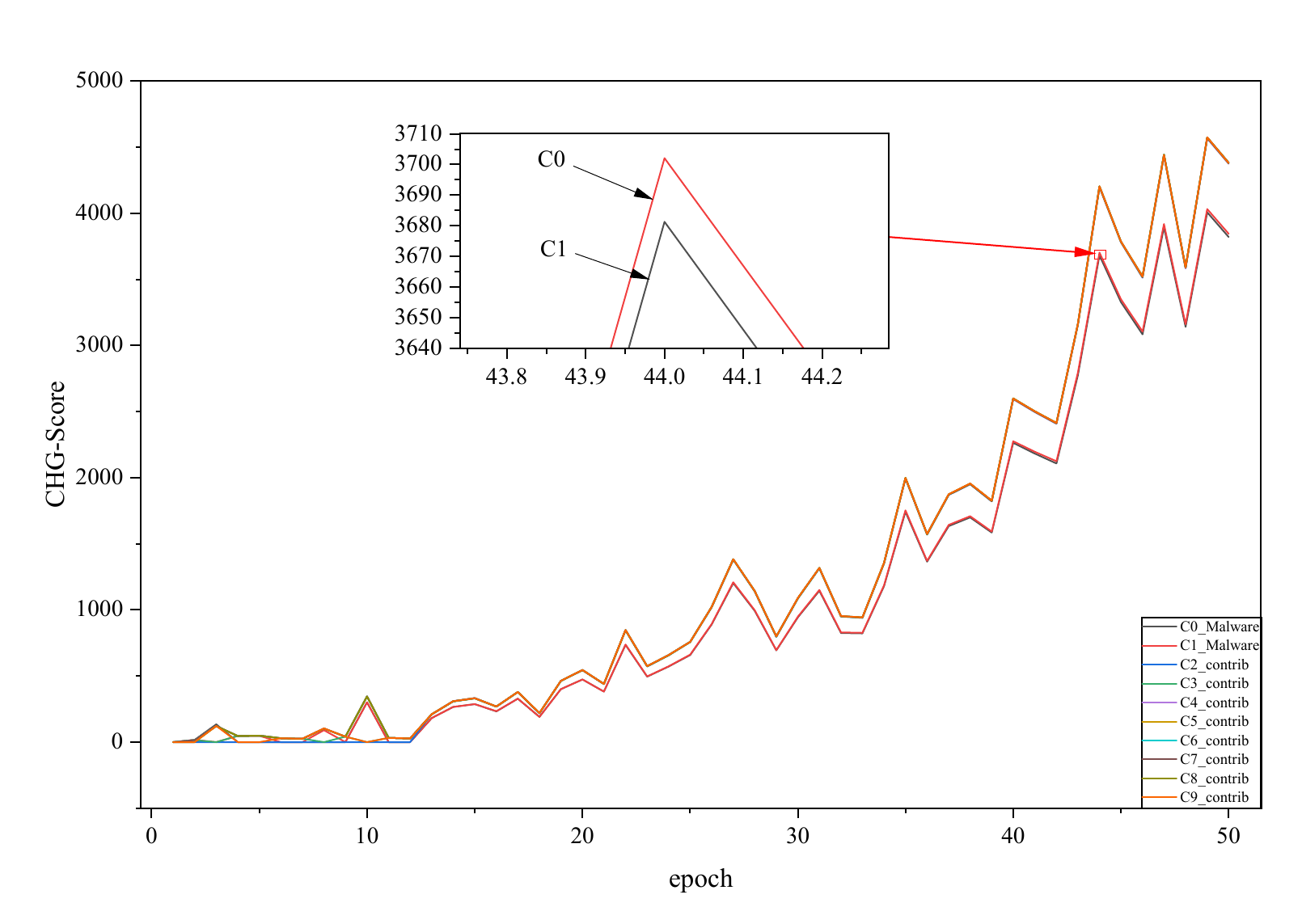}
    \caption{Trend of CHG contribution values over time for Secure\_CHG.}
    \label{fig:filter_chg_trend}
\end{figure}

\begin{figure}[H]
    \centering
    \includegraphics[width=\linewidth]{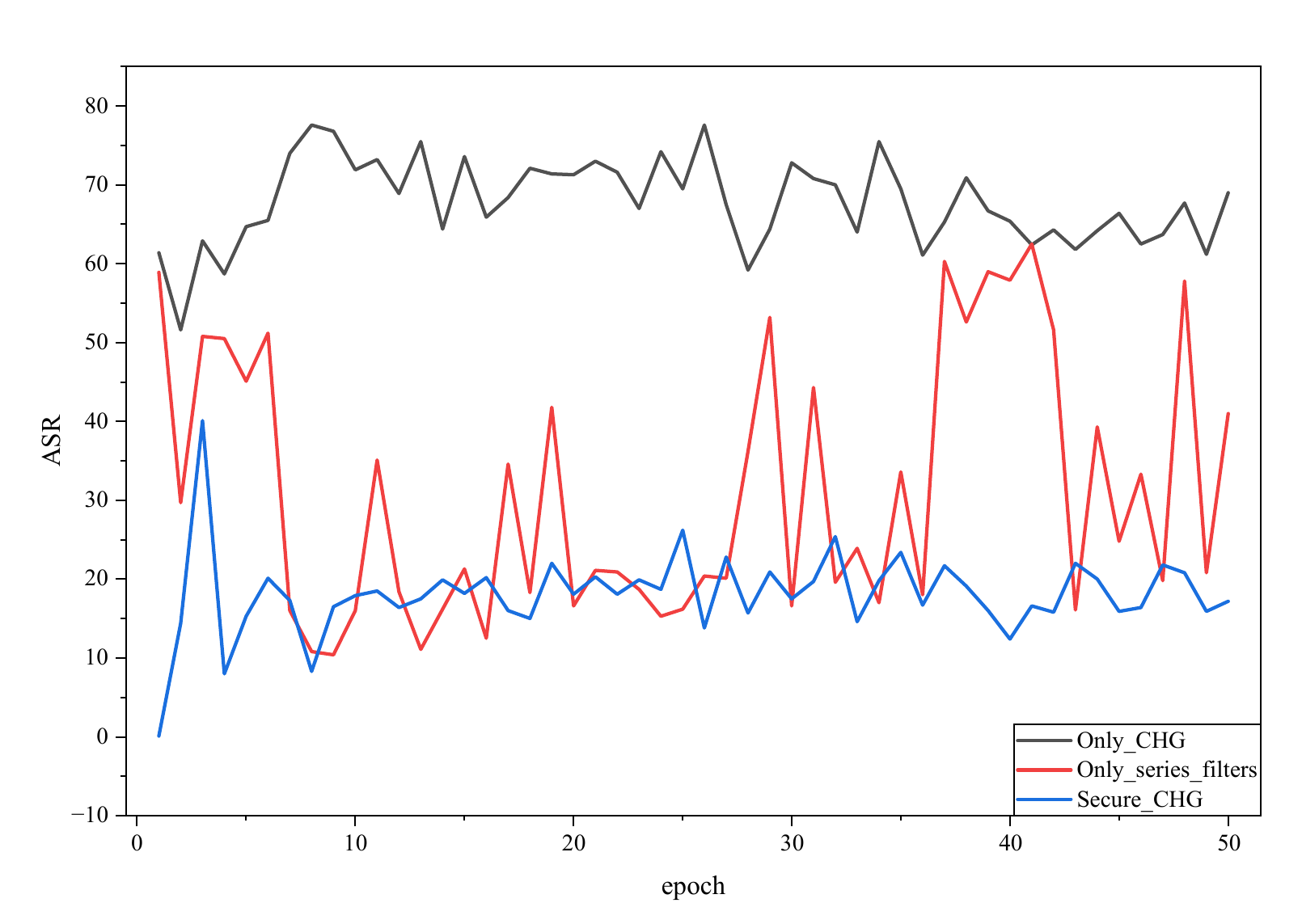}
    \caption{ASR trends for ablation study comparisons.}
    \label{fig:ablation_asr}
\end{figure}

First, relying solely on contribution evaluation (Only\_CHG) exposes a critical weakness during the early stages. As depicted in Fig.~\ref{fig:chg_only_trend}, without early constraints, malicious clients successfully hijack the initial training phase to accumulate erroneously high contribution scores, culminating in a 67.31\% ASR (Fig.~\ref{fig:ablation_asr}). Second, standalone cascaded filtering (Only\_series\_filters) exhibits severe late-stage ASR instability (33.06\% ASR), empirically confirming the ``Late-stage Failure'' dilemma.

In contrast, the proposed Secure-CHG framework dynamically orchestrates two orthogonal defenses to circumvent this dual dilemma, achieving a consistently low ASR of 18.84\%. A direct comparison between Fig.~\ref{fig:chg_only_trend} and Fig.~\ref{fig:filter_chg_trend} distinctly illuminates the mechanism behind this success. During early training, morphology-based statistical defenses (e.g., the FL-Defender filter in Section~\ref{subsec:micro_screening}) intercept anomalous updates. Crucially, this early-stage intervention effectively prevents attackers from accumulating unwarranted trust; consequently, the late-stage contribution scores of malicious nodes are fundamentally suppressed (Fig.~\ref{fig:filter_chg_trend}). Post-convergence, the CHG trust mechanism (Section~\ref{sec:chg_verification}) precisely isolates the remaining disguised adversaries via ``hardness'' divergence (malicious hardness remains high while benign approaches zero). Concurrently, Secure-CHG yields an Avg.\ All-Acc of 60.50\%, demonstrating that it deeply suppresses hard-to-detect adversaries while well-preserving global model utility.
\subsection{Generalization Verification}

To evaluate the generalization capability of the Secure-CHG framework, we conducted comparative experiments across three different dataset and model architecture scenarios:
\begin{enumerate}
    \item \textbf{MedMNIST} with MedMNIST-Net (Medical Image Classification).
    \item \textbf{CIFAR-10} with LeNet (Natural Image Classification).
    \item \textbf{NEU Surface Defect Database (NEU-SDDB)} with ResNet18 (Industrial Defect Detection), using bi-directional label flipping (Source $\leftrightarrow$ Target).
\end{enumerate}

The results are summarized in Table~\ref{tab:generalization}.

\begin{table*}[htbp]
    \centering
    \caption{Generalization Performance Across Different Datasets and Models}
    \label{tab:generalization}
    \resizebox{\textwidth}{!}{%
    \begin{tabular}{llccccccc}
        \toprule
        \multicolumn{2}{l}{\textbf{Dataset / Model}} & \textbf{FedAvg} & \textbf{FedMedian} & \textbf{TrimmedMean} & \textbf{MultiKrum} & \textbf{FLTrust} & \textbf{Krum\_CHG} & \textbf{Secure-CHG (ours)} \\ 
        \midrule
        \multirow{3}{*}{\shortstack[l]{MedMNIST /\\ MedMNIST-Net}} 
        & Acc & 77.66 & 73.05 & 73.96 & 78.70 & 77.36 & 77.72 & \textbf{78.91} \\
        & $Src_{Acc}$ & 32.84 & 35.47 & 33.46 & 56.12 & 37.25 & 57.04 & \textbf{60.93} \\
        & ASR & 9.47 & 10.47 & 10.43 & 9.44 & 10.49 & 10.38 & \textbf{8.88} \\ 
        \midrule
        \multirow{3}{*}{\shortstack[l]{CIFAR-10 /\\ LeNet}} 
        & Acc & 60.45 & 60.45 & 62.35 & 61.14 & \textbf{62.40} & 61.25 & 60.50 \\
        & $Src_{Acc}$ & 19.52 & 28.03 & 31.28 & 27.56 & 19.91 & 48.20 & \textbf{49.87} \\
        & ASR & 47.12 & 34.82 & 37.80 & 41.97 & 50.84 & 20.82 & \textbf{18.84} \\ 
        \midrule
        \multirow{3}{*}{\shortstack[l]{NEU-SDDB /\\ ResNet18}} 
        & Acc & 35.80 & \textbf{56.02} & 50.56 & 36.37 & 39.24 & 28.72 & 52.44 \\
        & $Src_{Acc}$ & 37.25 & 83.28 & \textbf{86.81} & 82.21 & 27.16 & 81.37 & 69.80 \\
        & ASR & 40.83 & 31.52 & 42.75 & 46.00 & 58.85 & 53.89 & \textbf{24.88} \\ 
        \bottomrule
    \end{tabular}%
    }
\end{table*}

Results show that Secure-CHG consistently achieves best or near-best performance in ASR. Specifically, in the MedMNIST task, it achieved the lowest ASR (8.88\%) and highest accuracy. In the CIFAR-10 scenario, its ASR (18.84\%) substantially outperforms traditional methods like MultiKrum. 

Notably, on the challenging NEU-SDDB dataset with bi-directional flipping, the framework suppressed ASR to 24.88\% while maintaining high main task accuracy, demonstrating strong adaptability to different distributions and data characteristics.

\subsection{Defense Against Advanced Threats: Backdoor Attacks}

To evaluate robustness against backdoor attacks, we simulate a scenario where malicious clients injected a $3 \times 3$ pixel trigger into training data and mislabeled them to a specific target class. We compared FedAvg, Secure-CHG, TrimmedMean, RFA, and FoolsGold over 80 rounds.

\begin{figure}[htbp]
    \centering
    \includegraphics[width=0.8\linewidth]{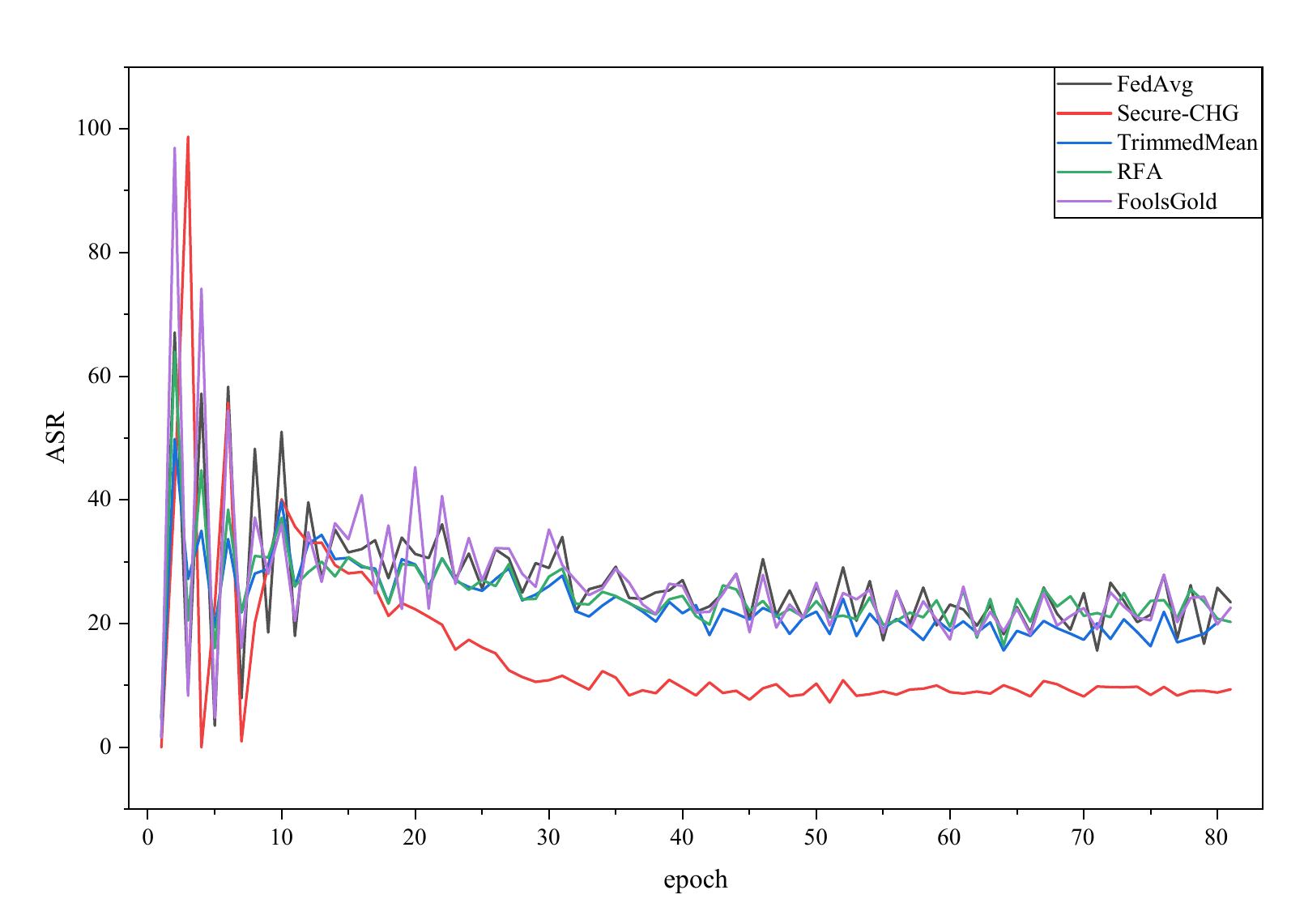}
    \caption{ASR trend under Backdoor Attack.}
    \label{fig:backdoor_asr}
\end{figure}

\begin{figure}[htbp]
    \centering
    \includegraphics[width=0.8\linewidth]{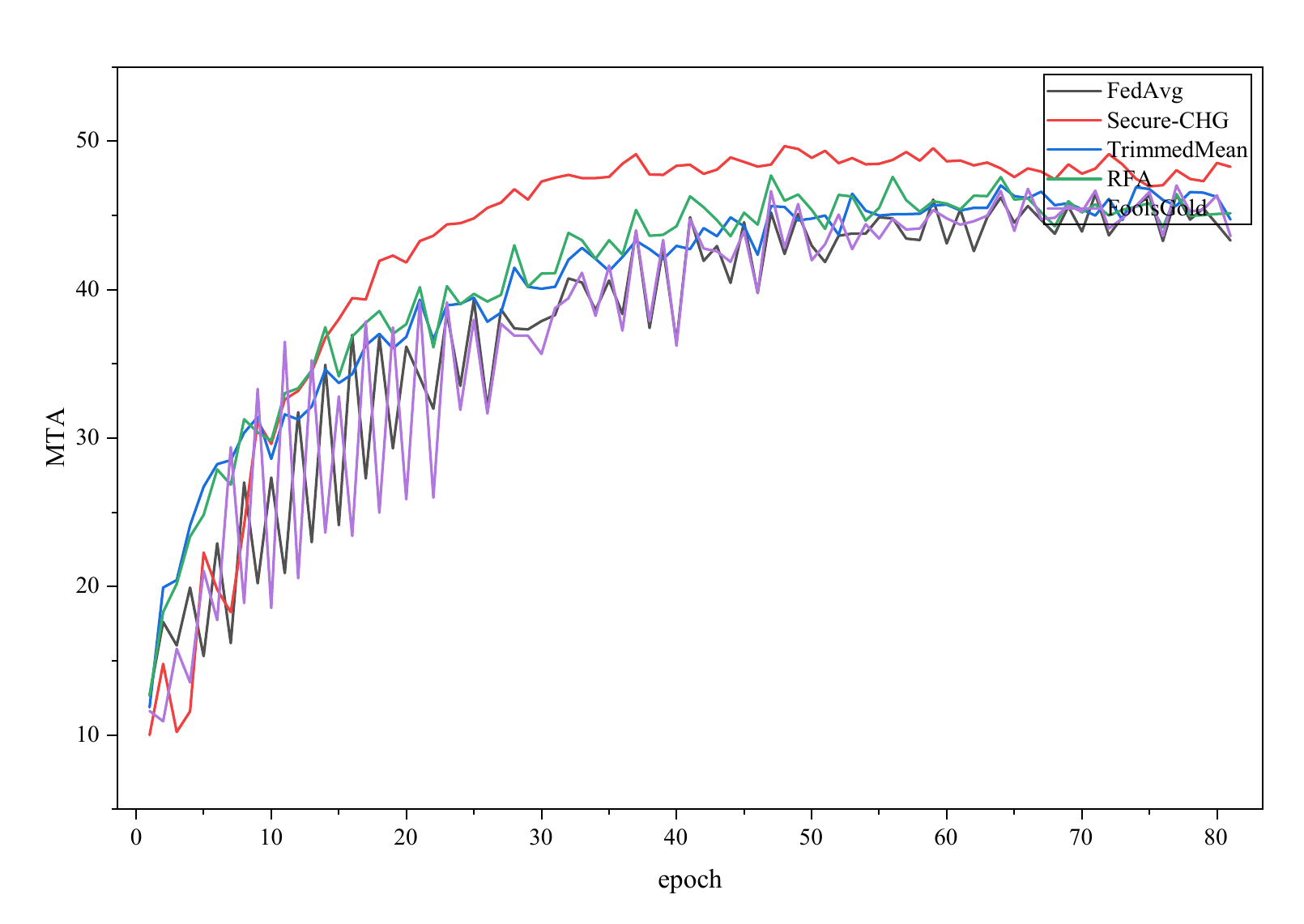}
    \caption{MTA trend under Backdoor Attack.}
    \label{fig:backdoor_mta}
\end{figure}

\begin{table}[htbp]
    \centering
    \caption{Robustness Against Backdoor Attacks (Rounds 31-80 Avg.)}
    \label{tab:backdoor_results}
    \resizebox{\columnwidth}{!}{%
    \begin{tabular}{lccccc}
        \toprule
        \textbf{Metric} & \textbf{FedAvg} & \textbf{Secure-CHG (ours)} & \textbf{TrimmedMean} & \textbf{RFA} & \textbf{FoolsGold} \\
        \midrule
        \textbf{ASR (\%)} & 23.43 & \textbf{9.38} & 20.27 & 22.60 & 23.01 \\
        \textbf{MTA (\%)} & 43.29 & \textbf{48.28} & 44.72 & 45.14 & 43.59 \\
        \bottomrule
    \end{tabular}%
    }
\end{table}

Experimental results (Table~\ref{tab:backdoor_results}, Fig.~\ref{fig:backdoor_asr}, Fig.~\ref{fig:backdoor_mta}) demonstrate Secure-CHG's superior defense. It suppresses backdoor ASR to a very low level of 9.38\%, substantially outperforming TrimmedMean (20.27\%) and FoolsGold (23.01\%), while maintaining a higher MTA(Main Task Accuracy) of 48.28\%. This demonstrates that the hybrid defense mechanism is robust against stealthy backdoor attacks, as the CHG-Shapley mechanism accurately captures persistent anomalies exposed by the conflict between the poisoning task and the main task.

\section{Conclusion and Future Works}
\label{sec:conclusion}

\subsection{Conclusion}
Federated Learning (FL) currently faces a ``dual dilemma'' regarding security and fairness in real-world deployment. The first challenge is the \textit{Late-stage Failure} of security mechanisms, where traditional defenses relying on statistical morphology (e.g., Euclidean distance, Cosine similarity) inevitably collapse as model updates converge and benign/malicious gradients homogenize. The second is the \textit{Shapley Value Paradox} in fairness assessment, where the theoretical gold standard becomes impractical due to exponential computational complexity ($O(2^n)$) and reliance on external proxy data.

To systematically address these challenges, this paper proposes \textbf{Secure-CHG}. The key innovation lies in a paradigm shift from ``Morphological Detection'' based on superficial statistics to ``Contribution Verification'' based on intrinsic value. We show that even when geometric features collapse, malicious clients maintain a high ``Hardness'' due to the conflict between poisoning tasks and the main task, making them distinguishable in the contribution dimension.

Key contributions of this framework include:
\begin{itemize}
    \item \textbf{Dual-Layer Hybrid Defense Architecture:} We design a dynamic defense lifecycle that utilizes statistical filtering to mitigate early-stage volatility and adaptively switches to contribution verification upon convergence to resolve late-stage failure.
    \item \textbf{Efficient CHG-Shapley Evaluation:} By deriving a closed-form solution for the ``Hardness-Gradient'' utility function, we reduce computational complexity from exponential to constant time ($O(1)$), eliminating the need for model retraining or proxy data.
    \item \textbf{Endogenous Governance Loop:} The framework unifies defense, evaluation, and incentives, transforming contribution scores into dynamic aggregation weights, forming a self-purifying ecosystem.
\end{itemize}

Extensive experiments demonstrate the superior robustness of Secure-CHG. Against label-flipping attacks on CIFAR-10, it maintains a low Attack Success Rate (ASR) of 18.84\%, significantly outperforming the MultiKrum baseline (41.97\%).Furthermore, against stealthy Backdoor Attacks, it suppresses ASR to 9.38\%. Generalization tests on MedMNIST and NEU-SDDB further confirm its general effectiveness in overcoming late-stage failure.

\subsection{Future Works}
Although Secure-CHG demonstrates significant advantages, optimizing for extreme data heterogeneity remains a key challenge. Future research within Non-IID environments will primarily focus on three directions: First, introducing \textbf{distribution-aware hardness calibration} (e.g., via lightweight global priors or clustering) to distinguish ``honest hard samples'' from malicious anomalies, thereby preserving fairness for long-tail data holders. Second, extending the framework to \textbf{Personalized Federated Learning (pFL)} by adapting CHG-Shapley to assess marginal contributions to specific communities or personalized layers, constructing a more inclusive trust network that rewards unique local contributions. Finally, exploring \textbf{privacy-enhanced contribution verification} through the integration of Zero-Knowledge Proofs (ZKP) or Secure Multi-Party Computation (SMPC), enabling clients to prove the validity of their contributions without exposing raw distribution data under strict privacy constraints.

\bibliographystyle{IEEEtran}
\bibliography{references}

\begin{IEEEbiography}[
  {\raisebox{0.2in}{\includegraphics[width=1.15in, height=1.4in, clip,keepaspectratio]{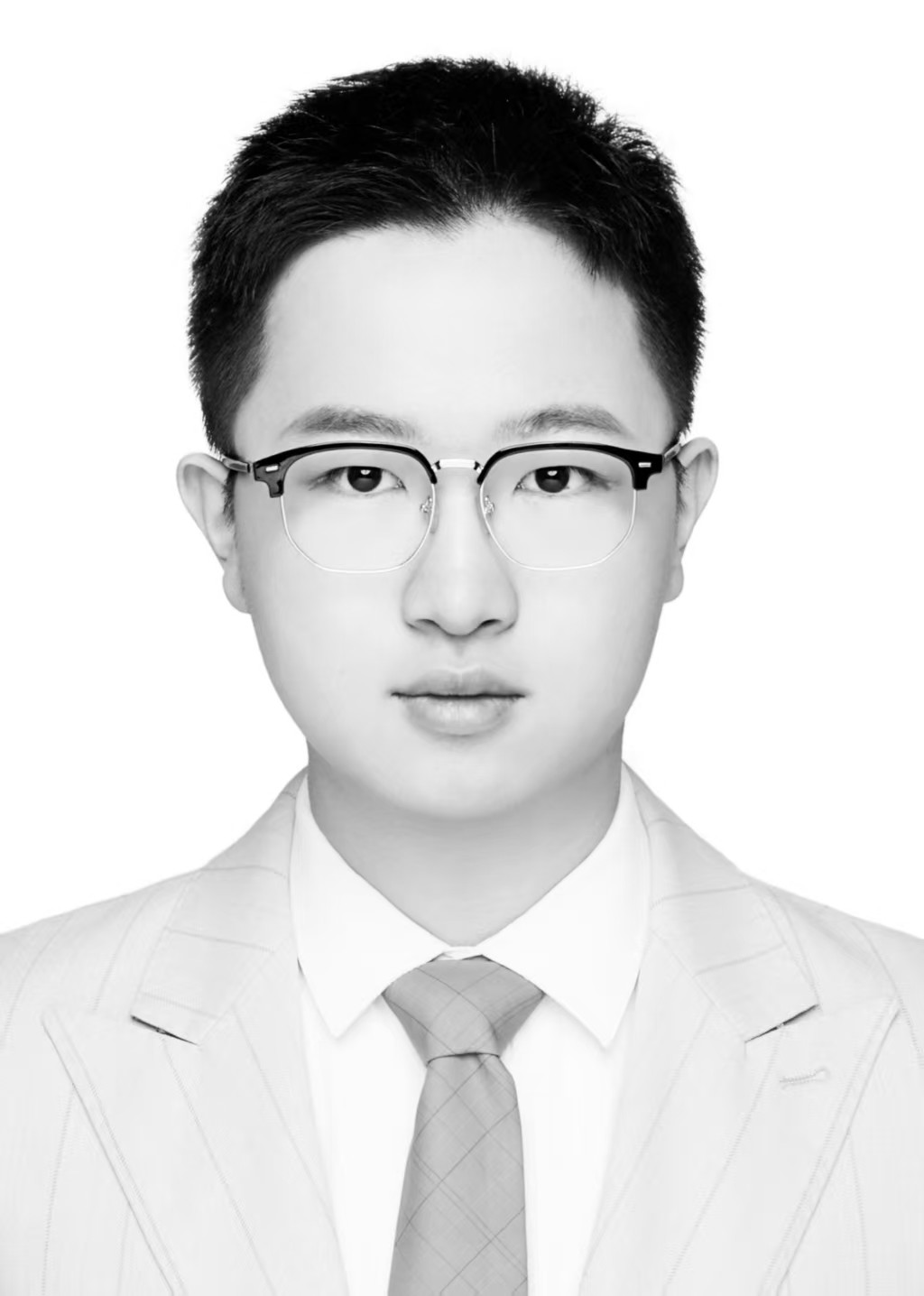}}}
]{Guanming Che}
is currently pursuing a B.S. degree in information security with the Software College, Northeastern University, Shenyang, China. His research interests include network security and federated learning.
\end{IEEEbiography}
\begin{IEEEbiography}[
  {\raisebox{0.15in}{\includegraphics[width=1in, height=1.25in, clip,keepaspectratio]{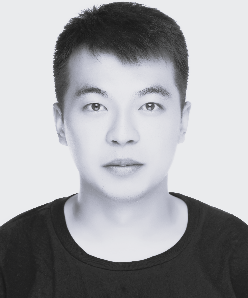}}}
]{Qiang Wang}
received the M.S. degree in information security and the Ph.D. degree in software engineering from Northeastern University, Shenyang, China, in 2016 and 2020, respectively. He is currently an associate Professor at Northeastern University, Shenyang, China. His research interests include verifiable computation and secure multi-party computation.
\end{IEEEbiography}
\begin{IEEEbiography}[
  {\raisebox{0.05in}{\includegraphics[width=1in, height=1.25in, clip,keepaspectratio]{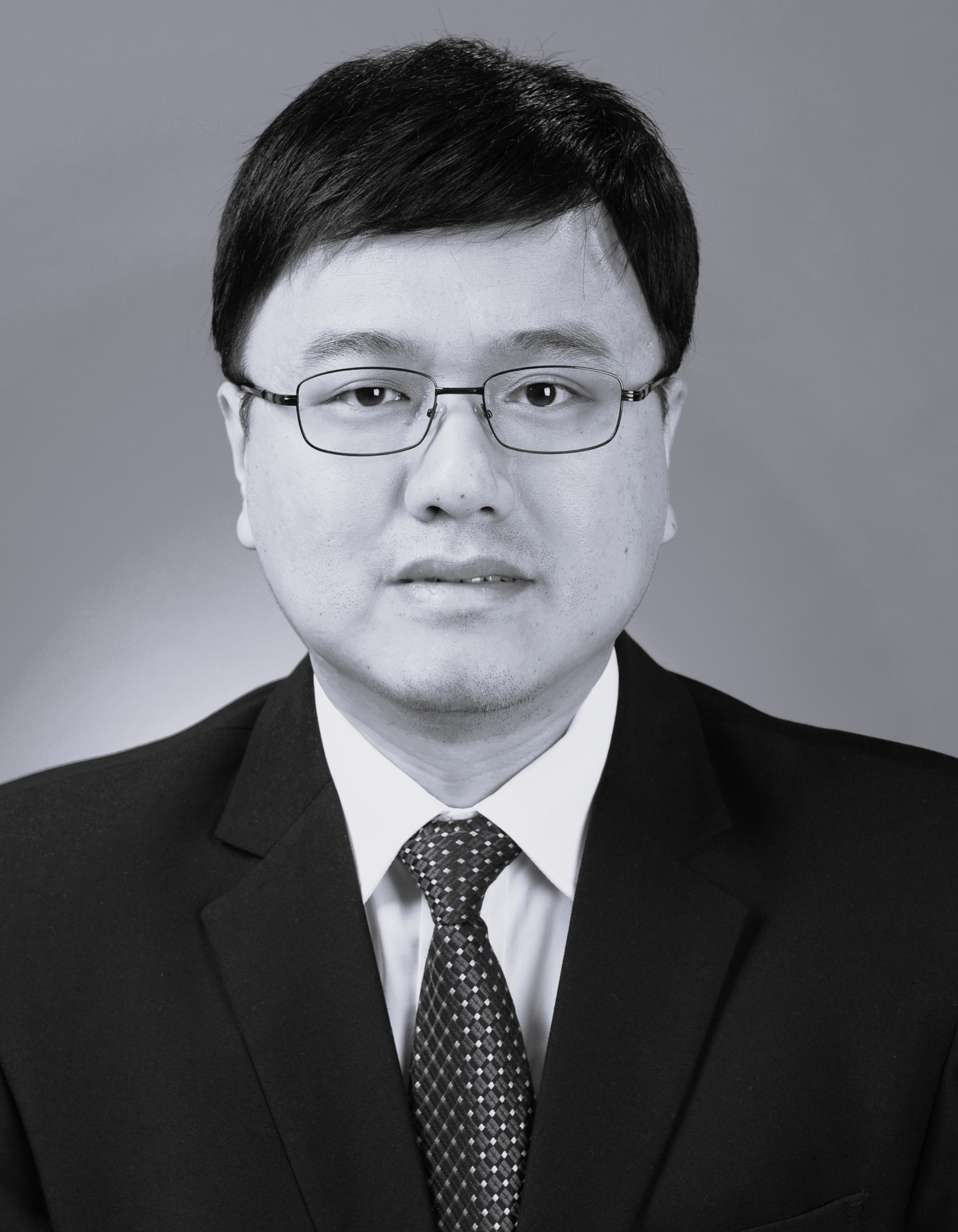}}}
]{Jian Xu} received the Ph.D. degree in computer application technology from Northeastern University, Shenyang, China, in 2013. He is currently a Professor and Doctoral Supervisor of Software College in Northeastern University. His research interests include cryptography and network security.
\end{IEEEbiography}
\begin{IEEEbiography}[
  {\raisebox{0.1in}{\includegraphics[width=1in, height=1.25in, clip,keepaspectratio]{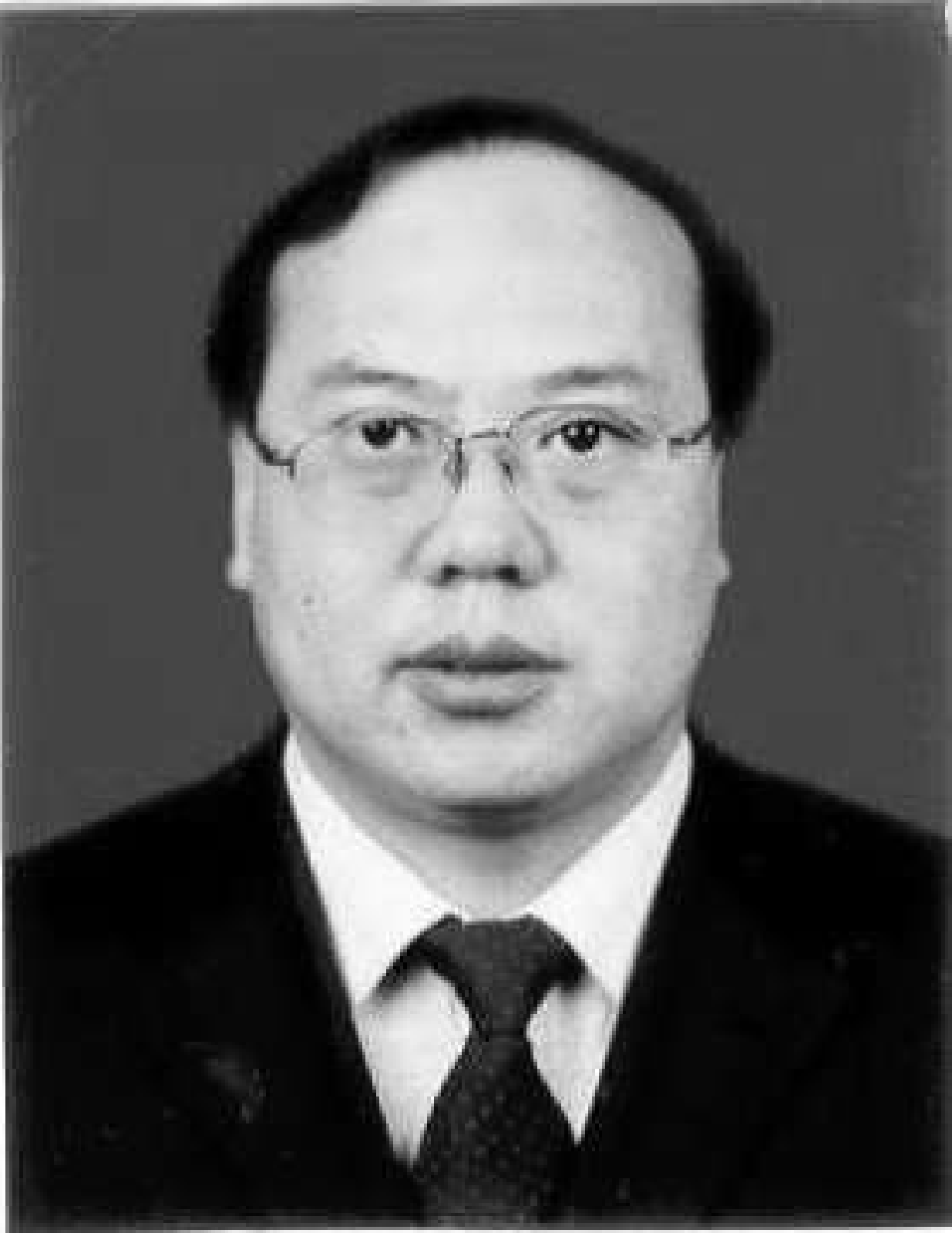}}}
]{Fucai Zhou} received the Ph.D. degree of Computer Software and Theory at Northeastern University, Shenyang, China, in 2001. He is currently a Professor and Doctoral Supervisor of Software College in Northeastern University. His research interests include cryptography, network security, trusted computing, secure cloud storage, software security, basic theory and critical technology in electronic commerce.
\end{IEEEbiography}
\end{document}